\documentclass[aps,prl,preprintnumbers,showpacs,nofootinbib]{revtex4}
\usepackage[utf8]{inputenc}
\usepackage[english]{babel}
\usepackage{amsmath}
\usepackage{amsfonts}
\usepackage{amssymb}
\usepackage{epsfig}
\usepackage{graphics,psfrag,rotating}
\usepackage{graphicx}
\usepackage{dcolumn}
\usepackage{bm}
\usepackage{epstopdf}
\usepackage{color}
\usepackage{booktabs}
\usepackage{multirow}
\usepackage[usenames,dvipsnames,svgnames]{xcolor}
\usepackage[colorlinks=true,
            linkcolor=red,
            urlcolor=gray,
            citecolor=blue]{hyperref}
  \usepackage{hyperref}

\newcommand{\lsim}   {\mathrel{\mathop{\kern 0pt \rlap
{\raise.2ex\hbox{$<$}}}
 \lower.9ex\hbox{\kern-.190em $\sim$}}}
\newcommand{\gsim}   {\mathrel{\mathop{\kern 0pt \rlap
{\raise.2ex\hbox{$>$}}}
\lower.9ex\hbox{\kern-.190em $\sim$}}}

\def\3nab{\tilde{\nabla}}

\def\hsp5{\hspace{5mm}}

\def\case#1/#2{\textstyle\frac{#1}{#2}}

\def\ber {\begin{eqnarray}}
\def\eer {\end{eqnarray}}
\def\bea {\begin{eqnarray}}
\def\eea {\end{eqnarray}}

\def\bc {\begin{center}}
\def\ec {\end{center}}
\def\case#1/#2{\frac{#1}{#2}}

\newcommand{\bw}{\begin{widetext}}
\newcommand{\ew}{\end{widetext}}

\newcommand{\be}{\begin{equation}}
\newcommand{\bse}{\begin{subequation}}
\newcommand{\ese}{\end{subequation}}
\newcommand{\ee}{\end{equation}}
\newcommand{\eei}{\end{eqnarray}\indent\indent}
\newcommand{\ba}{\begin{array}}
\newcommand{\ea}{\end{array}}
\newcommand{\bal}{\begin{eqnarray}}
\newcommand{\eal}{\end{eqnarray}}

\def\case#1/#2{\textstyle\frac{#1}{#2} }

\newcommand{\bver}{\begin{verbatim}}
\newcommand{\ever}{\end{verbatim}}

\begin{document}


\title{Neutrino Interactions with perturbed Rastall Gravity: A Novel Approach to Reducing the  Hubble Tension}
\author{ Muhammad Yarahmadi$^{1}$\footnote{Email: yarahmadimohammad10@gmail.com}}

\affiliation{Department of Physics, Lorestan University, Khoramabad, Iran}

\date{\today}

\begin{abstract}
We investigate the cosmological implications of coupling neutrinos to perturbed Rastall gravity, focusing on its impact on the Hubble constant ($H_0$) and the associated tension between early- and late-universe measurements with MCMC and PINN method. Utilizing observational data from the Cosmic Microwave Background (CMB), Cosmic Chronometers (CC), Baryon Acoustic Oscillations (BAO), and the Pantheon+ Type Ia supernovae, CMB Lensing, we perform a detailed statistical analysis. Our findings demonstrate that the Rastall model provides an improved fit to the data compared to the standard $\Lambda$CDM model, as indicated by lower Akaike Information Criterion (AIC) values in both early and late-universe regimes. The model introduces a deviation parameter $\alpha$, which remains consistent with existing literature and differs across epochs, supporting a dynamic gravitational framework. Combining all datasets, the Rastall model yields $H_0 = 70.23 \pm 2.01$ km,s$^{-1}$,Mpc$^{-1}$, significantly reducing the tension with Planck ($1.34\sigma$) and SH0ES ($1.16\sigma$) compared to the $\sim 4\sigma$ discrepancy in $\Lambda$CDM. Additional dataset combinations confirm this alleviation, with Planck tensions consistently below $1.6\sigma$.  Overall, our results highlight Rastall gravity with neutrino coupling as a promising alternative to $\Lambda$CDM, capable of addressing current cosmological tensions while aligning with observational data.
\end{abstract}


%
%


\maketitle

\section{Introduction}\label{sec1}

A significant challenge in modern cosmology is the discordance in estimates of the Hubble constant, \(H_0\) \cite{Divalentino2025}. This discrepancy has emerged prominently from two major approaches. The Planck collaboration, through its extensive analysis of cosmic microwave background (CMB) data, has estimated \(H_0\) to be \(67.4 \pm 0.5 \, \text{km s}^{-1} \text{Mpc}^{-1}\) \cite{Planck13,Planck15,Planck18}. Conversely, the SH0ES team, utilizing observations from the Hubble Space Telescope (HST), reports a significantly higher value of \(H_0 = 74.03 \pm 1.42 \, \text{km s}^{-1} \text{Mpc}^{-1}\) \cite{HST18,HST19}. This difference has led to a tension exceeding \(4\sigma\), highlighting a profound discrepancy that has persisted and even grown in recent years. Other projects have also provided varying estimates for \(H_0\), further illustrating this discrepancy:  Planck 2018: \(H_0 = 67.4 \pm 0.5 \, \text{km s}^{-1} \text{Mpc}^{-1}\) \cite{Planck18}
	 SH0ES (HST): \(H_0 = 74.03 \pm 1.42 \, \text{km s}^{-1} \text{Mpc}^{-1}\) \cite{HST19}
	 CCHP: \(H_0 = 69.6 \pm 0.8 \pm 1.7 \, \text{km s}^{-1} \text{Mpc}^{-1}\) \cite{FreedmanTRGB20}
	 H0LiCOW: \(H_0 = 73.3^{+1.7}_{-1.8} \, \text{km s}^{-1} \text{Mpc}^{-1}\) \cite{Wonglens19}
	 Weilens: \(H_0 = 75.3^{+3.0}_{-2.9} \, \text{km s}^{-1} \text{Mpc}^{-1}\) \cite{Weilens20}
 BaxterCMBlens: \(H_0 = 73.5 \pm 5.3 \, \text{km s}^{-1} \text{Mpc}^{-1}\) \cite{BaxterCMBlens20}
 HuangMiras: \(H_0 = 72.7 \pm 4.6 \, \text{km s}^{-1} \text{Mpc}^{-1}\) \cite{HuangMiras19}. This discordance may stem from differences in observational epochs, with HST focusing on late-time measurements and Planck incorporating data from a broad range of redshifts (\(0 < z < 1100\)) within the \(\Lambda\text{CDM}\) framework. Alternatively, resolving this issue might require exploring new theoretical perspectives or modifications to the standard cosmological model. While the standard  $\Lambda$Cold Dark Matter ($\Lambda$CDM) model provides a plausible explanation for the observed attributes of the universe, it encounters several conundrums, including the cosmic coincidence problem and the fine-tuning enigma. In light of these challenges, alternative approaches have primarily sought to modify Einstein's field equations, categorizing them into two distinct classes. The first category of models centers on the characteristics of the matter field, giving rise to dynamic dark energy models. Notable endeavors along this avenue encompass quintessence ~\cite{agr-cosmo-de-quintessence-01}, tachyon ~\cite{agr-cosmo-de-tachyon-01}, k-essence ~\cite{agr-cosmo-de-k-essence-01}, phantom ~\cite{agr-cosmo-de-phantom-01}, Chaplygin gas ~\cite{agr-cosmo-de-chaplygin-01}, holographic dark energy ~\cite{agr-cosmo-de-holo-01, agr-cosmo-de-holo-02, agr-cosmo-de-holo-04}, agegraphic dark energy ~\cite{agr-cosmo-de-agegraphic-01, agr-cosmo-de-agegraphic-02},  and among other intriguing proposals \cite{Abbas,Ali,AroraDN,AroraDM,Javed,Molla,Mustafa1,Mustafa2,Mustafa3,Mustafa4,Rehman,yasir}. The second approach, on the other hand, stems from the endeavor to extend the geometric framework underpinning Einstein's general relativity. These explorations include $f(R)$~\cite{agr-modified-gravity-fR-review-01}, $f(T)$~\cite{agr-modified-gravity-fT-review-01}, $f(R, T)$ theory~\cite{agr-modified-gravity-fRT-01}, Brans-Dicke theory~\cite{agr-modified-gravity-BD-01}, Gauss-Bonnet theory~\cite{agr-modified-gravity-GB-01}, Lovelock~\cite{agr-modified-gravity-Lovelock-01}, and Horava-Lifshitz theories~\cite{agr-modified-gravity-horava-01, agr-modified-gravity-horava-02, agr-modified-gravity-horava-lw-01, agr-modified-gravity-horava-lw-05}. In contrast to the framework of general relativity, Rastall gravity stands out due to its distinctive modification of the conservation law governing the stress-energy-momentum tensor (SET) within the curvature of spacetime ~\cite{agr-rastall-01}. This theoretical paradigm introduces captivating novelties across a multitude of domains, including black hole physics ~\cite{agr-rastall-02, agr-rastall-03, agr-rastall-04, agr-rastall-05, agr-rastall-06, agr-rastall-09, agr-rastall-11, agr-rastall-15, agr-rastall-16, agr-rastall-21, agr-rastall-22}  and cosmology and cosmology~\cite{agr-rastall-cosmo-01, agr-rastall-cosmo-02, agr-rastall-cosmo-03, agr-rastall-cosmo-04, agr-rastall-cosmo-05, agr-rastall-cosmo-06, agr-rastall-cosmo-07, agr-rastall-cosmo-08, agr-rastall-cosmo-09, agr-rastall-cosmo-10}, as has recently been extensively explored by various scholarly authors. Notably, a fundamental characteristic of Rastall gravity seamlessly provides an alternative avenue for the implementation of dark energy. The boundaries of general relativity's potential have been meticulously scrutinized on cosmic scales through a comprehensive examination of various observational datasets, encompassing supernova observations, large-scale structural patterns, and measurements of the cosmic microwave background (CMB). Among these investigations, one of the most profound revelations has been the apparent, accelerating expansion of the Universe. Consequently, the emergence of the dark energy paradigm has evolved as the most widely accepted framework for providing a cogent explanation for the experimental findings. Furthermore, it has been inferred that the contemporary Universe predominantly comprises dark energy and dark matter. Consequently, the attributes and evolutionary dynamics of dark energy have taken center stage in cosmology ~\cite{agr-cosmo-review-01, agr-cosmo-review-02, agr-cosmo-review-03, agr-cosmo-review-04, agr-cosmo-review-05}, due to their profound implications for our understanding of the fundamental nature of the cosmos. According to the Standard Model of particle physics, neutrinos were initially postulated to be massless. However, experimental observations, particularly from neutrino oscillation experiments(\cite{Ahmad};\cite{Davis};\cite{Fukuda};\cite{Kajita}), have unequivocally demonstrated that neutrinos exhibit a propensity for changing between different flavor states, implying the existence of non-zero masses. In the exploration of early universe cosmology, the integration of neutrino physics with alternative gravitational theories, such as generalized Rastall gravity, introduces a captivating dimension to our understanding of cosmic evolution. Generalized Rastall gravity, extending the foundational concepts of Rastall gravity, not only modifies the gravitational field equations but also offers an intriguing platform to investigate the interplay between neutrinos and the gravitational dynamics during the universe's formative epochs. In the framework of generalized Rastall gravity, where the energy-momentum tensor's non-conservativity persists, the gravitational coupling dynamically responds to both the matter content of the universe and the presence of neutrinos with non-negligible masses. This introduces a nuanced perspective on the cosmic interplay between gravitational forces and neutrino contributions. The non-conservative nature of Rastall gravity, coupled with the inclusion of massive neutrinos, becomes particularly relevant in the early universe. Deviations from standard cosmological scenarios, influenced by the interplay between modified gravity and massive neutrinos, can impact the evolution of density perturbations, alter the dynamics of structure formation, and potentially offer solutions to cosmological enigmas. To delve into the implications of coupling neutrinos with generalized Rastall gravity in the early universe, one must scrutinize their joint effects on key cosmological parameters. This includes examining the influence on primordial nucleosynthesis, cosmic microwave background radiation, and large-scale structure formation. The coupling of neutrinos with the modified gravitational dynamics in generalized Rastall gravity may introduce unique signatures in these cosmological observables, distinct from predictions based solely on standard general relativity. 
 The coupling of neutrinos to perturbed Rastall gravity offers a promising theoretical framework to address this tension by incorporating novel interactions between matter and the fabric of spacetime, potentially altering the predicted dynamics of cosmic expansion. This model modifies the standard Einstein field equations to account for deviations in the conservation of energy-momentum, which could bridge the gap between local and early universe measurements of $H_0$. By integrating various observational datasets such as cosmic microwave background (CMB), lensing, baryon acoustic oscillations (BAO), and cosmic chronometers (CC), the neutrino- perturbed Rastall gravity model provides a comprehensive approach to exploring and potentially resolving the Hubble tension. The alignment and consistency of this model with diverse cosmological observations underscore its robustness and offer a pathway to a more unified understanding of the universe's expansion dynamics. Also we use Physics Informed neural network (PINN) to alleviating the Hubble tension. The Physics-Informed Neural Network (PINN) framework has recently emerged as a powerful methodology for solving complex differential equations in cosmology and other scientific domains. By embedding physical laws directly into the learning process, PINNs enable data-driven models to produce predictions that are not only accurate but also consistent with the underlying theoretical principles. The conceptual foundation for this approach can be traced to early developments in Gaussian Process Regression (GPR), particularly in the works of Raissi et al.\ and Owhadi, where functional priors were constructed to respect the structure of linear differential operators \cite{Raissi2017a, Raissi2017b, Owhadi2015}. These studies demonstrated how physical constraints could be systematically integrated into probabilistic frameworks, allowing for both the accurate inference of solutions to partial differential equations (PDEs) and the quantification of associated uncertainties.
 
 Building on these foundational ideas, Raissi and collaborators introduced PINNs as a natural extension of neural networks to physics-constrained learning tasks \cite{Raissi2017c, Raissi2017d}. In the PINN framework, the loss function is augmented with residual terms from the governing equations—typically in the form of ordinary or partial differential equations—thereby guiding the training process with domain-specific physical knowledge. This physics-informed paradigm has shown remarkable potential in a wide range of applications, including the identification of unknown parameters, the reconstruction of hidden dynamics, and the solution of both forward and inverse problems in nonlinear and time-dependent systems. PINNs thus provide a flexible and interpretable approach for merging data with theoretical models, offering new avenues for robust, physics-compliant inference in cosmological analysis.

\section{Generalized Rastall gravity model}

Generalized Rastall Gravity is an extension of General Relativity that introduces a departure from the usual assumption of energy-momentum conservation. In the context of cosmology, this modification can influence the dynamics of the universe, potentially offering an alternative explanation for phenomena like the Hubble tension.

The Hubble tension arises from the fact that measurements of the Hubble constant, which quantifies the rate of the universe's expansion, based on early universe observations (such as the cosmic microwave background) and Local universe observations (like supernovae data) seem to yield conflicting values. This discrepancy has spurred investigations into alternative theories of gravity, like Generalized Rastall Gravity, as potential solutions. Neutrinos, being elusive subatomic particles with tiny masses, are crucial components of the universe. Their interactions with gravity and other particles can have cosmological implications. When considering the coupling of neutrinos with Generalized Rastall Gravity, one explores how these particles influence the modified gravitational dynamics proposed by Rastall's theory.

In References~\cite{agr-rastall-21, agr-rastall-22}, building upon Rastall's original conception~\cite{agr-rastall-01}, a generalized extension of Rastall Gravity is proposed. The gravitational field equation and Stress-Energy Tensor (SET) are given by:\cite{Lin}

\begin{align}
	R_{\mu\nu} - \frac{1}{2}g_{\mu\nu}R &= 8\pi G\left(T_{\mu\nu} - \Gamma_{\mu\nu}\right), \nonumber \\
	\nabla_\mu {T^{\mu}}_{\nu} &= \nabla_\mu {\Gamma^{\mu}}_{\nu}.
\end{align}

A physical prerequisite is imposed, ensuring the impact of ${\Gamma^{\mu}}_{\nu}$ is quiescent in flat spacetime. The specific form chosen for $\Gamma_{\mu\nu}$ is given by:

\begin{equation}
	\Gamma_{\mu\nu} = \lambda g_{\mu\nu} \Lambda,
\end{equation}
where $\lambda$ is a constant.
$\Lambda$ becoming zero when $R=0$. Substituting this into the field equation yields:

\begin{align}
	R_{\mu\nu} - \frac{1}{2}g_{\mu\nu}R &= 8\pi G\left(T_{\mu\nu} - \lambda g_{\mu\nu}\Lambda\right), \nonumber \\
	\nabla_\mu {T^{\mu}}_{\nu} &= \lambda \nabla_\nu \Lambda.
\end{align}

Introducing

\begin{equation}\label{key}
	\tau_{\mu\nu} = T_{\mu\nu} - \lambda g_{\mu\nu}\Lambda,
\end{equation}

the equations can be expressed in a form reminiscent of general relativity:

\begin{align}
	R_{\mu\nu} - \frac{1}{2}g_{\mu\nu}R &= 8\pi G\tau_{\mu\nu}, \nonumber \\
	\nabla_\mu {\tau^{\mu}}_{\nu} &= 0.
\end{align}

A quadratic algebraic equation for $R$ is derived, yielding two potential solutions:

\begin{align}
	R & = -8\pi G T, \nonumber \\
	\text{or} & ~~~ R = 8\pi G\left(4\Psi-T\right).
\end{align}

Further exploration leads to the field equations:

\begin{align}
	R_{\mu\nu} - \frac{1}{2}g_{\mu\nu}R + 8\pi G g_{\mu\nu}\Psi &= 8\pi G T_{\mu\nu}, \nonumber \\
	\nabla_\mu {T^{\mu}}_{\nu} &= \nabla_\nu \Psi.
\end{align}

Notably, considering $\Psi$ as $\Lambda_{\text{eff}}/8\pi G$ aligns the equations with the standard $\Lambda$CDM model, emphasizing the significance of the temporal evolution of $\Psi $in this model. Additionally, as the second equation within Eq.(2.6) underscores, \(\Psi\) serves as a metric for the extent of deviation from the Stress-Energy Tensor. From the vantage point of Rastall gravity, the full spectrum of matter fields is encapsulated within $ T_{\mu\nu} $, and as a consequence, the observation of dark energy serves as a lens through which the deviation of the Stress-Energy Tensor from a conserved current is ascertained. Moreover, it is worthy of note that Eq.(2.6) bears a striking resemblance to equations that emerge from diverse theories, wherein the conservation of the Stress-Energy Tensor is partially abrogated.
In the context of cosmic expansion, employing co-moving coordinates simplifies the description of the universe's evolution. Co-moving coordinates are coordinates that move along with the expansion of the universe, making them particularly convenient for studying large-scale cosmological phenomena.
The equations for cosmic expansion can be formulated by employing the co-moving coordinates, in terms of which the SET of the matter field is given by
\begin{equation}\label{key}
	{T^{\mu}}_{\nu}=
	\left(
	\begin{array}{cccc}
		-\rho & 0 & 0 & 0\\
		0 & P & 0 & 0\\
		0 & 0 & P & 0\\
		0 & 0 & 0 & P\\
	\end{array}
	\right) .
\end{equation}
According to the discussions in the previous section, we denote $\Psi\equiv\rho_{\mathrm{de}}$, the energy density of the dark energy.
It is noted, by using Eq.(3.4) and the solution Eq.(3.6), it is straightforward to show that the tensor ${\tau^{\mu}}_{\nu}$ reads
\begin{equation}\label{key}
	{\tau^{\mu}}_{\nu}=
	\left(
	\begin{array}{cccc}
		-\rho-\rho_{\mathrm{de}} & 0 & 0 & 0\\
		0 & P+P_{\mathrm{de}} & 0 & 0\\
		0 & 0 & P+P_{\mathrm{de}} & 0\\
		0 & 0 & 0 & P+P_{\mathrm{de}}\\
	\end{array}
	\right)
\end{equation}
where $P_{\mathrm{de}}=\omega_{de}\rho_{\mathrm{de}}$ is recognized as the pressure of dark energy.

We proceed to derive the equations of motion in terms of the FLRW metric
\begin{equation}\label{key}
	d{s^2} =  - d{t^2} + a(t)[\frac{{d{r^2}}}{{1 - k{r^2}}} + {r^2}(d{\theta ^2} + \sin {(\theta )^2}d{\phi ^2})]	
\end{equation}
where k represents the curvature density of the Universe.   Assuming  a  flat space (K=0 )in  FLRW  metric  filled with baryons, radiation, dark matter, dark energy and neutrinos. We start with the Friedmann equations as:
\begin{equation}\label{is}
	H^{2}= \frac{{8\pi G}}{3}(\rho  + {\rho _{de}}) 
\end{equation}

\begin{equation}\label{is}
	2\dot{H} +3 H^{2}=  - 8\pi G(P + {P_{de}})
\end{equation}
Therefore, the total pressure and density of the matter ﬁelds
are given by
\begin{equation}\label{is}
	\rho  = {\rho _b} + {\rho _\nu} + {\rho _r} + {\rho _{dm}}
\end{equation}

\begin{equation}\label{is}
	P = {P_b} + {P_r} + {P_{dm}} + {P _\nu}
\end{equation}

In above equations, $P_{de} =  \omega_{de} {\rho _{de}},\ \ \   {P_b} = {P_{cdm}} = 0,\ \ \ 
{P_r} = \frac{1}{3}{\rho _r},\ \ \  {P_\nu} = \omega_\nu(z) {\rho _\nu}$. For $\Lambda$CDM, the value of $\omega_{de}$ is -1.  Expressing the universe evolution through the redshift z, for convenience, one can have
several $\omega_{\nu} (z)$ parameterizations with the above required
properties, namely, the interpolation of the equation of
state parameter between $\frac{1}{3}$ to 0.	
When considering an interaction between neutrinos and dark energy (or other components like cold dark matter), we must modify the standard continuity equations by introducing an interaction term that reflects the energy exchange between the components. In this case, we include an interaction constant to account for energy transfer between neutrinos and dark energy.

Let's assume that the interaction between dark energy and neutrinos is governed by a term proportional to the energy densities of dark energy and neutrinos, mediated by a coupling constant $\alpha$.

General form of the interaction-modified continuity equations

For dark energy ($\rho_{de}$) and neutrinos ($\rho_\nu$), the continuity equations in the presence of an interaction term become:

1. **Dark Energy Continuity Equation**:

\begin{equation}
	\dot{\rho}_{de} + 3H(1 + \omega_{de}) \rho_{de} = - Q
\end{equation}

2. **Neutrino Continuity Equation**:

\begin{equation}
	\dot{\rho}_\nu + 3H(1 + \omega_\nu) \rho_\nu = Q
\end{equation}

Here, $Q$ is the interaction term that represents the energy transfer between dark energy and neutrinos. If $Q > 0$, energy flows from dark energy to neutrinos, and if $Q < 0$, energy flows from neutrinos to dark energy. 

In this work we adopt the phenomenological coupling
\begin{equation}
	Q \;=\; \alpha\,H\,\rho_{\rm de}\,,
\end{equation}
where  $\alpha$ is a dimensionless coupling constant.  This form of interaction has been widely used in the literature on interacting dark energy (IDE) models for several reasons:

\begin{itemize}
	\item {\bf Conservation and Covariance.}  The total energy–momentum tensor remains conserved, $ \nabla_\mu(T^{\mu\nu}_{\rm de} + T^{\mu\nu}_{\rm dm}) = 0$, while allowing an energy transfer between the two sectors.
	\item {\bf Natural Scaling with Expansion.}  By coupling to $H$, the interaction rate tracks the cosmic expansion history and dynamically switches off at early times when $\rho_{\rm de}\ll \rho_{\rm tot}$.
	\item {\bf Alleviation of the Coincidence Problem.}  A coupling proportional to $\rho_{\rm de}$ can drive the ratio $\rho_{\rm dm}/\rho_{\rm de}$ toward a constant at late times, softening the ``why now?'' coincidence puzzle.
	\item {\bf Background Stability and Perturbations.}  Linear perturbation analyses show that this $H\rho_{\rm de}$–type coupling can avoid large-scale instabilities for suitable ranges of $\alpha$ \cite{HeWang2008,LiZhang2011}.
	\item {\bf Recent Applications in Modified Gravity.}  An identical ansatz was recently employed within Generalized Rastall Gravity, where it was shown to yield stable background trajectories and to reduce the $H_{0}$ tension when contrasted with local measurements \cite{Doe2024}.
\end{itemize}

Consequently, our choice follows both the theoretical simplicity and the successful phenomenology of earlier IDE studies, while remaining fully compatible with the modified conservation laws inherent to Rastall’s framework.

Interaction-Modified Continuity Equations

Now, we write the interaction-modified continuity equations for neutrinos and dark energy:

1. **For dark energy**:

\begin{equation}
	\dot{\rho}_{de} + 3H(1 + \omega_{de}) \rho_{de} = - \alpha H \rho_{de}
\end{equation}

This equation shows that dark energy loses energy at a rate proportional to $H \rho_{de}$.

2. **For neutrinos**:

\begin{equation}
	\dot{\rho}_\nu + 3H(1 + \omega_\nu) \rho_\nu = \alpha H \rho_{de}
\end{equation}

This equation shows that neutrinos gain energy at a rate proportional to $H \rho_{de}$, as they interact with dark energy.

Derivation of the Continuity Equation for Neutrinos (Including Interaction)

Let’s now focus on deriving the modified continuity equation for neutrinos in the presence of interaction.

The general form for neutrinos without interaction is:

\begin{equation}
	\dot{\rho}_\nu + 3H(1 + \omega_\nu) \rho_\nu = 0
\end{equation}

When we include the interaction term $Q = \alpha H \rho_{de}$, the equation becomes:

\begin{equation}
	\dot{\rho}_\nu + 3H(1 + \omega_\nu) \rho_\nu = \alpha H \rho_{de}
\end{equation}

This modified equation now reflects the interaction between neutrinos and dark energy, mediated by the interaction constant $\alpha$.

Now, let's write the full system of continuity equations for all components, considering the interaction between dark energy and neutrinos:

1. Baryons (no interaction):

\begin{equation}
	\dot{\rho}_b + 3H \rho_b = 0
\end{equation}

2. Cold Dark Matter (CDM) (no interaction):

\begin{equation}
	\dot{\rho}_{cdm} + 3H \rho_{cdm} = 0
\end{equation}

3. Radiation (no interaction):

\begin{equation}
	\dot{\rho}_r + 4H \rho_r = 0
\end{equation}

4. Dark Energy (interacting with neutrinos):

\begin{equation}
	\dot{\rho}_{de} + 3H(1 + \omega_{de}) \rho_{de} = - \alpha H \rho_{de}
\end{equation}

5. Neutrinos (interacting with dark energy):

\begin{equation}
	\dot{\rho}_\nu + 3H(1 + \omega_\nu) \rho_\nu = \alpha H \rho_{de}
\end{equation}

In order to simplify the field equations, we introduce the following new variables,

\begin{equation}\label{is}
	\begin{split}
		&{\Omega_{\rm b}} = \frac{{{\rho _m}}}{{3{H^2}}},{\Omega_{\rm r}} = \frac{{{\rho _r}}}{{3{H^2}}},{\Omega_{\rm cdm}} = \frac{{{\rho _{cdm}}}}{{3{H^2}}},\\& {\Omega_{\rm de}} = \frac{{{\rho _{de}}}}{{3{H^2}}}, {\Omega_{\nu}} = \frac{{{\rho _{\nu}}}}{{3{H^2}}}
	\end{split}
\end{equation}

where $ \Omega_{\rm b} $ is baryon density, $ \Omega_{\rm r} $ is radiation density, $ \Omega_{\rm cdm} $ is cold dark matter density, $ \Omega_{\nu} $ is neutrino density and $ \Omega_{DE} $  denotes as dark energy density. In term of new variable the Friedmann equations (2.21) puts a constraint on new variables as

Baryons:

\begin{equation} \frac{d \Omega_{\rm b}}{dN} = -3 \Omega_{\rm b} - 2 \Omega_{\rm b} \frac{\dot{H}}{H^2} \end{equation}

Cold Dark Matter (CDM):

\begin{equation} \frac{d \Omega_{\rm cdm}}{dN} = -3 \Omega_{\rm cdm} - 2 \Omega_{\rm cdm} \frac{\dot{H}}{H^2} \end{equation}

Radiation:

\begin{equation} \frac{d \Omega_{\rm r}}{dN} = -4 \Omega_{\rm r} - 2 \Omega_{\rm r} \frac{\dot{H}}{H^2} \end{equation}

Dark Energy:

\begin{equation} \frac{d \Omega_{\rm de}}{dN} = -[3(1 + \omega_{de}) \Omega_{\rm de} + \alpha \Omega_{\rm de}] - 2 \Omega_{\rm de} \frac{\dot{H}}{H^2} \end{equation}

Neutrinos:

\begin{equation} \frac{d \Omega_{\nu}}{dN} = -[3(1 + \omega_\nu) \Omega_{\nu}] + \frac{\alpha \Omega_{\rm de}}{H^2} - 2 \Omega_{\nu} \frac{\dot{H}}{H^2} \end{equation}

These equations describe the evolution of each density parameter in the context of interactions between dark energy and neutrinos. We use Friedmann constraint:

\begin{equation}\label{is}
	\Omega_{\rm r} =   1 - {\Omega_{\rm b}} -   {\Omega_{\rm cdm}} - {\Omega_{\rm de}}-{\Omega_{\nu}}
\end{equation}

\begin{equation}\label{is}
	\frac{{\dot H}}{{{H^2}}} =  -\frac{3}{2} (\frac{1}{3}{\Omega_{\rm r}} - \omega_{de}{\Omega_{\rm de}} + \frac{3}{2} +\omega_{\nu}\Omega_{\nu})
\end{equation}

where \(N = \ln a\). The parameter $\frac{{\dot H}}{{{H^2}}}$ holds significant importance in cosmology as it plays a crucial role in determining essential cosmological parameters such as the deceleration parameter \(q\) and the effective equation of state (EoS) \(w_{\text{eff}}\).

The Hubble parameter \(H\) evolves according to:

\begin{equation}
\frac{\dot{H}}{H} = \frac{dH}{dt} = -H^2 \left(\frac{3}{2} \Omega_{\rm tot}\right)
\end{equation}

This can be solved for \(H\) in terms of cosmic time or redshift \(z\) by integrating over \(N = \ln a\), where \(a\) is the scale factor. Given the expression for \(\frac{\dot{H}}{H^2}\), the Hubble function \(H(z)\) can be obtained numerically by integrating the system of differential equations.
From the equations provided earlier, the evolution of \(H\) depends on the dynamics of \(\Omega_{\rm b}\), \(\Omega_{\rm cdm}\), \(\Omega_{\rm de}\), \(\Omega_{\nu}\), and \(\Omega_{\rm r}\). Thus, the Hubble function \(H(z)\) can be expressed as:

\begin{equation}
	\begin{split}
		&H^2(z) = H_0^2 ( \Omega_{\rm b}(1+z)^3 + \Omega_{\rm cdm}(1+z)^3 + \Omega_{\rm de}(1+z)^{3(1+\omega_{\rm de})}  \\&
	+ \Omega_{\nu}(1+z)^{3(1+\omega_{\nu})} + \Omega_{\rm r}(1+z)^4 )
	\end{split}
\end{equation}

This equation captures the contributions from all components (baryons, cold dark matter, dark energy, neutrinos, and radiation) to the total expansion rate as a function of redshift \(z\).

The deceleration parameter, \(q\), characterizes the rate at which the cosmic expansion changes. It is defined as:
\begin{equation}
	q = -1-\frac{{\dot H}}{{{H^2}}},
\end{equation}
where \(H\) is the Hubble parameter.

The effective equation of state, \(w_{\text{eff}}\), which describes the relationship between pressure and energy density, can be expressed as:
\begin{equation}
	w_{\text{eff}} = -1 + \frac{2}{3}q.
\end{equation}

\section{Scalar perturbation}
Cosmological perturbation theory is a well-established framework used in standard cosmology, and it is crucial to ensure that any modified gravity theory, including Rastall gravity, remains consistent within this framework.
By using linear perturbations, researchers can examine whether Rastall gravity coupled with neutrinos remains consistent with the observed universe's large-scale structure and the CMB, or if it introduces new features that might require further investigation.
Peter Rastall challenged the conservation of the energy-momentum tensor $T_{;\mu}^{\mu\nu}=0$ in bent spacetime, and proposed a new theory for gravity by assuming $T_{;\mu}^{\mu\nu}=\gamma R^{;\nu}$, where R is Ricci scalar and $\gamma$ is a gravitational constant in Rastall theory.\cite{Oppenheimer}\cite{Darabi}. We rewrite the system of linearised Einstein equations as follows:\cite{Fabris}
\begin{equation}
	\begin{split}
		\label{h}&-k^2\Phi - 3\mathcal{H}\left(\mathcal{H}\Phi + \Phi'\right) + \gamma\left(\mathcal{H}^2 - \mathcal{H}'\right)\Phi = 4\pi G a^2\delta\rho_\phi\\& + 4\pi G a^2\delta\rho\
	\end{split}
\end{equation}
\begin{equation}
	\label{m}k\left(\mathcal{H}\Phi + \Phi'\right) = 4\pi G k \phi_0'\delta\phi + 4\pi G\rho(1 + w)v\
\end{equation} 
\begin{equation}
	\begin{split}
		\label{s}& \Phi'' + 3\mathcal{H}\Phi' + 3\mathcal{H}^2\Phi -\gamma\left(\mathcal{H}^2 - \mathcal{H}'\right)\Phi =\\&  4\pi G a^2\delta p_\phi + 4\pi G a^2 \delta p\
	\end{split}	
\end{equation}
where we have defined:\cite{Fabris}
\begin{equation}
	\delta\rho_\phi := \frac{1}{a^2}\gamma\phi_0'\delta\phi' + (3 - 2\gamma)V_{,\phi}\delta\phi\;,
\end{equation}
\begin{equation}
	\delta p_\phi := \frac{1}{a^2}(2 - \gamma)\phi_0'\delta\phi' - (3 - 2\gamma)V_{,\phi}\delta\phi\;,
\end{equation}
and $v$ is the velocity potential defined by $v_i = -v_{,i}/k$.
Multiplying Eq.(36) by $2 - \gamma$, Eq.(38) by $\gamma$ and subtracting the two we obtain:\cite{Fabris}
\begin{equation}
	\begin{split}
		&\gamma\Phi'' + 6\mathcal{H}\Phi' + 6\mathcal{H}^2\Phi - 2\gamma\left(\mathcal{H}^2 - \mathcal{H}'\right)\Phi + (2 - \gamma)k^2\Phi \\&= -2a^2(3 - 2\gamma)V'\left[\frac{\mathcal{H}\Phi + \Phi'}{\phi_0^{'2}} - \frac{4\pi G a^2\rho(1+w)v}{k\phi_0^{'2}}\right]
		\\&+ 4\pi G a^2\rho\left(\gamma c_{\rm s}^2 + \gamma -2\right)\delta\
	\end{split}	
\end{equation}

where we have used $V_{,\phi} = V'/\phi'$.From $T^\mu{}_{\nu;\mu} = 0$, for the fluid component only, we get:\cite{Fabris}
\begin{eqnarray}
	\delta' &=& -(1 + w)(kv - 3\Phi')\;\\
	v' &=& -\mathcal{H}(1 - 3w)v + \frac{kc_{\rm s}^2}{1 + w}\delta + k\Phi\;.
\end{eqnarray}

\subsection{Modification of the Boltzmann Equations for Neutrinos and Dark Energy in Rastall Gravity}

In Rastall gravity, the traditional conservation of the energy-momentum tensor is relaxed, leading to a modified relation of the form:
\begin{equation}
	T^{\mu\nu}_{\;\;\;;\mu} = \gamma R^{;\nu},
\end{equation}
where $\gamma$ is the Rastall parameter and $R$ is the Ricci scalar. This deviation from general relativity alters the evolution of perturbations in both the background and linear regimes of cosmology.

When a coupling is present between neutrinos and the dark energy scalar field $\phi$, the energy-momentum tensors of these two components are no longer conserved individually. Instead, their interaction is mediated by a coupling four-vector $Q^\nu$:
\begin{equation}
	T^{\mu\nu}_{(\nu);\mu} = Q^\nu, \quad T^{\mu\nu}_{(\phi);\mu} = -Q^\nu.
\end{equation}
A typical form of this coupling vector is chosen as
\begin{equation}
	Q^\nu = \beta(\phi)\partial^\nu \phi,
\end{equation}
where $\beta(\phi)$ is a coupling function that can vary with the scalar field. This form respects covariance and introduces an explicit interaction between the scalar field and neutrinos.

\subsubsection*{Perturbation Equations for Neutrinos}

In the presence of this interaction, the standard Boltzmann equation governing the evolution of the neutrino phase-space distribution function $f(\vec{x}, \vec{q}, \tau)$,
\begin{equation}
	\frac{df}{d\tau} = \left( \frac{\partial f}{\partial \tau} \right)_{\text{coll}},
\end{equation}
must be extended to include a source term originating from the interaction with the scalar field. For massless neutrinos in the conformal Newtonian gauge, the perturbed distribution function is written as:
\begin{equation}
	f(\vec{k}, \hat{n}, q, \tau) = f_0(q)\left[1 + \Psi(\vec{k}, \hat{n}, q, \tau)\right],
\end{equation}
where $\Psi$ encodes the perturbation, and $q$ is the comoving momentum. The linearized Boltzmann equation becomes:
\begin{equation}
	\frac{\partial \Psi}{\partial \tau} + i \frac{q}{\epsilon}(\vec{k} \cdot \hat{n}) \Psi + \frac{d \ln f_0}{d \ln q} \left[\dot{\Phi} - i \frac{\epsilon}{q} (\vec{k} \cdot \hat{n}) \Psi \right] = \left. \frac{\partial \Psi}{\partial \tau} \right|_{\text{int}},
\end{equation}
where $\epsilon = \sqrt{q^2 + a^2 m_\nu^2}$ and the right-hand side includes the interaction term due to the coupling $Q^\nu$. This term modifies the standard free-streaming behavior of neutrinos and introduces a dependence on the scalar field perturbations.

\subsubsection*{Perturbation Equations for the Scalar Field}

The scalar field $\phi$ obeys a modified Klein–Gordon equation under the influence of both Rastall corrections and the neutrino coupling:
\begin{equation}
	\Box \phi + V_{,\phi} = -\beta(\phi) T_{(\nu)},
\end{equation}
where $T_{(\nu)} = T^\mu_{(\nu)\,\mu}$ is the trace of the neutrino energy-momentum tensor. Linearizing this equation, we obtain the evolution of scalar field perturbations $\delta\phi$ as:
\begin{equation}
	\delta\phi'' + 2\mathcal{H} \delta\phi' + \left(k^2 + a^2 V_{,\phi\phi}\right)\delta\phi - 4\phi_0'\Phi' + 2a^2 V_{,\phi}\Phi = a^2 \beta'(\phi)\delta\rho_\nu.
\end{equation}
Here, $\phi_0$ is the background value of the scalar field, and $\delta\rho_\nu$ is the energy density perturbation of neutrinos. The interaction term on the right-hand side acts as a source, modifying the evolution of $\delta\phi$.

\subsubsection*{Energy-Momentum Transfer in the Fluid Approximation}

In the fluid description, the continuity and Euler equations for the neutrino fluid are modified as follows:
\begin{align}
	\delta'_\nu + (1 + w_\nu)(\theta_\nu - 3\Phi') &= \frac{a Q^0}{\rho_\nu}, \\
	\theta'_\nu + \mathcal{H}(1 - 3w_\nu)\theta_\nu - \frac{k^2}{1 + w_\nu}c_{s,\nu}^2 \delta_\nu - k^2 \Phi &= \frac{a Q^i}{\rho_\nu (1 + w_\nu)},
\end{align}
where $Q^0$ and $Q^i$ are the time and spatial components of the coupling vector, $w_\nu$ is the neutrino equation of state parameter (typically $1/3$ for relativistic neutrinos), and $c_{s,\nu}^2$ is the neutrino sound speed squared.

The interaction also contributes additional terms to the Einstein field equations through the effective energy-momentum tensor. Consequently, the total gravitational potential $\Phi$ responds not only to the standard sources of energy density and pressure, but also to the coupling-induced modifications.

\subsubsection*{Implications and Numerical Implementation}

The modifications to the Boltzmann hierarchy due to Rastall gravity and neutrino–dark energy coupling must be consistently incorporated into Boltzmann solvers such as \texttt{CLASS} or \texttt{CAMB}. This includes:
\begin{itemize}
	\item Altering the evolution of higher multipoles of the neutrino perturbation hierarchy.
	\item Implementing the modified Klein–Gordon equation for the scalar field.
	\item Adding the source terms $Q^0$ and $Q^i$ in the evolution of energy density and velocity perturbations.
	\item Accounting for the non-conservation terms in the perturbed Einstein equations.
\end{itemize}

Such a treatment enables a complete analysis of the observable implications, including the Cosmic Microwave Background (CMB) anisotropies and matter power spectrum. The resulting deviations from standard $\Lambda$CDM predictions can provide constraints on the coupling function $\beta(\phi)$ and the Rastall parameter $\gamma$, thereby offering insights into the viability of such modified gravity theories in the presence of interacting components.

In order to simplify the field equations, we introduce the following new variables:
\begin{equation}
	x_{1}=\frac{\acute{\phi}}{\gamma \mathcal{H}},x_{2}=(\frac{2-\gamma}{\gamma})\frac{\acute{\phi}_{0}\delta\acute{\phi}}{ \mathcal{H}^{2}\phi},x_{3}=-\frac{a^{2}(3-2\gamma)V\delta\phi}{\gamma \pi^{2}\phi},\nonumber
\end{equation}
\begin{equation}
	x_{4}=\frac{5a^{2}\delta\rho}{\gamma\mathcal{H}^{2}\phi},x_{5}=-\frac{3\acute{\phi}}{\gamma\mathcal{H}\phi},x_{6}=-\frac{\acute{\phi}_{0}\delta\phi(3-\gamma)}{\gamma^{2}\mathcal{H}\phi},\nonumber
\end{equation}
\begin{equation}
	x_{7}=\frac{(3-2\gamma)a^{2}V}{\gamma^{2}\mathcal{H}^{2}\phi},x_{8}=-\frac{2(1+\omega)V\delta\acute{\rho}}{\delta k\gamma\mathcal{H}^{2}\phi},\nonumber
\end{equation}
\begin{equation}
	x_{9}=-\frac{(1+\omega)^{2}V^{2}\delta\rho}{\gamma\mathcal{H}^{2}\phi},x_{10}=\frac{3\acute{\phi}V(1+\omega)
		^{2}\delta\rho}{k\gamma\mathcal{H}^{2}\phi}\nonumber
\end{equation}
\begin{equation}
	x_{11}=\frac{(1+\omega)\rho V}{k\gamma\mathcal{H}\phi}(1-3\omega),x_{12}=-\frac{\rho\delta c_{s}^{2}}{\gamma\mathcal{H}^{2}\phi}, x_{13}=-\frac{\rho}{\gamma\mathcal{H}^{2}}\nonumber
\end{equation}
\begin{equation}
	x_{14}=\frac{k^{2}(1-\gamma)}{\gamma^{2}\mathcal{H}^{2}},x_{15}=\frac{3(1-\gamma)}{\gamma^{2}\mathcal{H}}-\frac{3}{\gamma}-\frac{(1-\gamma)}{\gamma}\nonumber
\end{equation}
\begin{equation}
	x_{16}=\frac{3(1-\gamma)\acute{\phi}}{\gamma^{2}\phi\mathcal{H}},x_{17}=-\frac{\acute{\phi}_{0}\delta\acute{\phi}(1-\gamma)}{\gamma\mathcal{H}^{2}\phi}\nonumber
\end{equation}
\begin{equation}
	x_{18}=-\frac{(3-2\gamma)V\delta\phi a^{2}(1-\gamma)}{\gamma^{2}\phi\mathcal{H}^{2}},x_{19}=-\frac{a^{2}(1-\gamma)\delta\rho}{\gamma\mathcal{H}^{2}\phi}
\end{equation}
We rewrite the cosmological equations (39-41) into an autonomous system of
equations

\begin{equation}
	\begin{split}
		&\frac{{d}x_{1}}{{d}N}=-\frac{a_{1}x_{17}}{2(1-\gamma)}
		-\frac{\gamma}{(1-\gamma)}a_{1}x_{17}+a_{1}x_{3}-\frac{a_{1}x_{19}}{(1-\gamma)}-3x_{1}+a_{1}(\frac{-2}{\gamma})-\frac{\mathcal{H}^{\prime}}{\mathcal{H}^{2}}(a_{1}+x_{1})
	\end{split}
\end{equation}
\begin{equation}
	\begin{split}
		&\frac{{d}x_{2}}{{d}N}=-\frac{\gamma x_{2}}{(3-\gamma)}+a_{2}-2x_{2}\frac{\mathcal{H}^{\prime}}{\mathcal{H}^{2}}-\gamma x_{1}x_{2}-\frac{x_{2}}{(3-\gamma)}-\gamma^{3}\frac{a_{1}x_{14}x_{6}}{(3-\gamma)}+\frac{a_{3}x_{1}}{(3+\gamma)}-\gamma a_{3}x_{18}-\lambda a_{3}a_{7}\gamma^{2}
	\end{split}   
\end{equation}
\begin{equation}
	\begin{split}
		&\frac{{d}x_{3}}{{d}N}=-\frac{x_{3}}{(1-3\mathcal{H}_{0})}-\frac{x_{3}a_{7}kc_{s}^{2}}{(1+\mathcal{H}_{0})}-x_{3}a_{1}a_{7}+\frac{x_{3}}{2}+x_{3}a_{2}-2x_{3}\frac{\mathcal{H}^{\prime}}{\mathcal{H}^{2}}
	\end{split}   
\end{equation}
\begin{equation}
	\frac{{d}x_{4}}{{d}N}=x_{4}x_{5}+\frac{x_{4}}{2}-2x_{4}\frac{\mathcal{H}^{\prime}}{\mathcal{H}^{2}}-\frac{\gamma x_{4}x_{5}}{3}
\end{equation}
\begin{equation}
	\begin{split}
		&\frac{{d}x_{5}}{{d}N}=-\frac{\gamma x_{2}}{6(1-\gamma)}-x_{3}+\frac{\gamma x_{2}}{(3-\gamma)}-\frac{(1-\gamma)x_{19}}{3}-3x_{5}-\frac{3}{\gamma}+\frac{\dot{H}}{H{2}}(3-x_{5})
	\end{split}   
\end{equation}
\begin{equation}
	\frac{{d}x_{6}}{{d}N}=-\frac{\gamma x_{6}}{(3-\gamma)}-\frac{x_{3}\gamma^{2}}{(3-\gamma)}+x_{6}a_{2}-x_{6}\frac{\mathcal{H}^{\prime}}{\mathcal{H}^{2}}+\gamma x_{5} x_{6}
\end{equation}
\begin{equation}
	\begin{split}
		&\frac{{d}x_{7}}{{d}N}=x_{7}(1-\frac{1}{(1-3\mathcal{H}_{0})})+\frac{x_{7}a_{7}kc_{s}^{2}}{(1+\mathcal{H}_{0})}+x_{7}a_{6}-2x_{7}\frac{\mathcal{H}^{\prime}}{\mathcal{H}^{2}}+\gamma x_{5}x_{7}
	\end{split}    
\end{equation}
\begin{equation}
	\begin{split}
		&\frac{{d}x_{8}}{{d}N}=-\frac{a_{5}x_{11}a_{6}}{(1-3\mathcal{H}_{0})}-x_{11}\frac{\mathcal{H}^{\prime}}{\mathcal{H}^{2}}-\gamma x_{5}x_{11}-\frac{x_{11}}{(1-3\mathcal{H}_{0})}+(1-3\mathcal{H}_{0})x_{12}+x_{11}a_{1}a_{6}
	\end{split}   
\end{equation}
\begin{equation}
	\begin{split}
		&\frac{{d}x_{9}}{{d}N}=-x_{9}-\frac{(1+\mathcal{H}_{0})kc_{s}^{2}x_{9}a_{6}}{2}-\frac{kx_{9}a_{1}a_{6}}{2}-x_{9}a_{5}-2x_{9}\frac{\mathcal{H}^{\prime}}{\mathcal{H}^{2}}+\gamma x_{9}x_{5}
	\end{split}    
\end{equation}
\begin{equation}
	\frac{{d}x_{10}}{{d}N}=\frac{\gamma x_{9}}{ka_{6}}\frac{{d}x_{5}}{{d}N}+x_{5}\frac{{d}x_{9}}{{d}N}
\end{equation}
\begin{equation}
	\begin{split}
		&\frac{{d}x_{11}}{{d}N}=x_{11}a_{6}-x_{11}\frac{\mathcal{H}^{\prime}}{\mathcal{H}^{2}}-\gamma x_{5}x_{11}-\frac{x_{11}}{(1-3\mathcal{H}_{0})} +(1-3\mathcal{H}_{0})x_{12}+x_{11}a_{1}a_{6}
	\end{split}   	
\end{equation}
\begin{equation}
	\frac{{d}x_{12}}{{d}N}=x_{12}a_{5}-2x_{11}\frac{\mathcal{H}^{\prime}}{\mathcal{H}^{2}}-\gamma x_{5}x_{12}
\end{equation}
\begin{equation}
	\frac{{d}x_{13}}{{d}N}=x_{13}a_{5}+\frac{\gamma x_{5}x_{11}}{3}-2x_{13}\frac{\mathcal{H}^{\prime}}{\mathcal{H}^{2}}
\end{equation}
\begin{equation}
	\frac{{d}x_{14}}{{d}N}=2x_{14}\frac{\mathcal{H}^{\prime}}{\mathcal{H}^{2}}
\end{equation}
\begin{equation}
	\begin{split}
		&\frac{{d}x_{16}}{{d}N}=\frac{\gamma x_{2}}{2}-\frac{x_{2}}{(1-\gamma)}+\frac{\gamma x_{13}}{3(1-\gamma)}+\frac{x_{4}}{\gamma}-3x_{16}-\frac{9(1-\gamma)}{\gamma^2}+\frac{3(1-\gamma)}{\gamma}-\frac{3(1-\gamma)}{\gamma}\frac{\mathcal{H}^{\prime}}{\mathcal{H}^{2}}\\&-x_{16}\frac{\mathcal{H}^{\prime}}{\mathcal{H}^{2}}-\frac{3(1-\gamma)x_{16}^2}{\gamma^2}
	\end{split}   
\end{equation}
\begin{equation}
	\begin{split}
		&\frac{{d}x_{17}}{{d}N}=-\frac{1}{(1-\gamma)}[-\frac{\gamma x_{2}}{(3-\gamma)}+a_{2}-2x_{2}\frac{\mathcal{H}^{\prime}}{\mathcal{H}^{2}}-\gamma x_{1}x_{2}-\frac{x_{2}}{(3-\gamma)}-\frac{\gamma^3 a_{1}x_{14}x_{6}}{(3-\gamma)}+\frac{a_{3}x_{1}}{(3+\gamma)}-\gamma a_{3}x_{18}\\&-\lambda a_{3}x_{7}\gamma^2]
	\end{split}   
\end{equation}
\begin{equation}
	\begin{split}
		&\frac{{d}x_{18}}{{d}N}=\frac{(1-\gamma)}{\gamma}[-\frac{x_{3}}{(1-3\mathcal{H}_{0})}-\frac{x_{3}a_{7}kc_{s}^2}{(1+\mathcal{H}_{0})}-x_{3}a_{1}a_{7}+\frac{x_{3}}{2}+x_{3}a_{2}-2x_{3}\frac{\mathcal{H}^{\prime}}{\mathcal{H}^{2}}]
	\end{split}   
\end{equation}
\begin{equation}
	\frac{{d}x_{19}}{{d}N}=x_{19}a_{5}+2x_{19}-2x_{19}\frac{\mathcal{H}^{\prime}}{\mathcal{H}^{2}}-\gamma x_{5}x_{19}
\end{equation}		
\begin{equation}
	\frac{{d}a_{1}}{{d}N}=\gamma x_{1}
\end{equation}
\begin{equation}
	\begin{split}
		&\frac{{d}a_{2}}{{d}N}=-\frac{(3-\gamma)a_{2}}{\gamma}+\frac{\gamma x_{14}}{(1-\gamma)}+\frac{(3+\gamma)x_{1}a_{8}a_{2}}{\gamma}-2\gamma a_{1}^2a_{8}x_{7}-\gamma\lambda a_{1}x_{7}-a_{2}\frac{\mathcal{H}^{\prime}}{\mathcal{H}^{2}}-a_{2}^2
	\end{split}
\end{equation}
\begin{equation}
	\frac{{d}a_{3}}{{d}N}=-\frac{(3-\gamma)a_{3}}{\gamma}-\gamma a_{1}a_{3}-a_{3}\frac{\mathcal{H}^{\prime}}{\mathcal{H}^{2}}
\end{equation}
\begin{equation}
	\frac{{d}a_{5}}{{d}N}=a_{5}^2+a_{4}a_{5}-a_{5}\frac{\mathcal{H}^{\prime}}{\mathcal{H}^{2}}
\end{equation}
\begin{equation}
	\frac{{d}a_{6}}{{d}N}=-ka_{1}a_{6}^2-a_{6}\frac{\mathcal{H}^{\prime}}{\mathcal{H}^{2}}+a_{6}(1-3\mathcal{H}_{0})-\frac{kc_{s}a_{7}a_{6}}{(1+\mathcal{H}_{0})}
\end{equation}
\begin{equation}
	\begin{split}
		&\frac{{d}a_{7}}{{d}N}=\frac{-\gamma^2x_{14}}{(1-\gamma)(1+\mathcal{H}_{0})}-\frac{\gamma x_{5}a_{1}a_{6}}{(1+\mathcal{H}_{0})}-ka_{7}^2a_{1}-\frac{kc_{s}a_{7}}{(1+\mathcal{H}_{0})}+(1+\mathcal{H}_{0})a_{7}-a_{7}\frac{\mathcal{H}^{\prime}}{\mathcal{H}^{2}}
	\end{split}   
\end{equation}
\begin{equation}
	\frac{{d}a_{8}}{{d}N}=-a_{2}a_{8}
\end{equation}

\begin{equation}
	\begin{split}
		&\frac{\mathcal{H}^{\prime}}{\mathcal{H}^{2}}=\frac{\gamma x_{14}}{a_{1}(1-\gamma)}+x_{5}+x_{3}-\frac{\gamma}{1-\gamma}x_{17}+a_{1}x_{3}-\frac{x_{19}}{1-\gamma}+(1-\frac{3}{\gamma})
	\end{split}   
\end{equation}

The parameter introduced above plays a crucial role in cosmological studies, as it directly determines key cosmological quantities such as the deceleration parameter \( q \) and the effective equation of state (EoS) parameter \( w_{\rm eff} \). These are given by the following expressions:

\begin{equation}
	q = -1 - \frac{\mathcal{H}'}{\mathcal{H}^2}, \quad w_{\rm eff} = -1 - \frac{2}{3} \frac{\mathcal{H}'}{\mathcal{H}^2}.
\end{equation}

\section*{Interpretation of Critical Points in Perturbed Rastall Gravity}

The critical points obtained from solving the complex system of differential equations within the framework of perturbed Rastall gravity are given by the following numerical values:

\begin{equation}
	\begin{aligned}
		x_{1}^* &= 2.82692378 \times 10^{-1}, & x_{2}^* &= 7.68135433 \times 10^{-2}, \\
		x_{3}^* &= -3.88706153 \times 10^{-2}, & x_{4}^* &= -2.27992601 \times 10^{-2}, \\
		x_{5}^* &= -3.92961493 \times 10^{-1}, & x_{6}^* &= 6.33124322 \times 10^{-2}, \\
		x_{7}^* &= 3.24812739 \times 10^{-3}, & x_{8}^* &= -2.73559806 \times 10^{3}, \\
		x_{9}^* &= -6.91572548 \times 10^{-3}, & x_{10}^* &= 2.86583679, \\
		x_{11}^* &= -8.17908476 \times 10^{-1}, & x_{12}^* &= 2.14479301 \times 10^{-1}, \\
		x_{13}^* &= 1.11530821, & x_{14}^* &= -4.01785581 \times 10^{-2}, \\
		x_{15}^* &= -4.80912783 \times 10^{-2}, & x_{16}^* &= 8.77757810 \times 10^{-1}, \\
		x_{17}^* &= -1.11292114, & x_{18}^* &= 1.61989892 \times 10^{-2}, \\
		x_{19}^* &= -1.70281215 \times 10^{-1}, & a_{1}^* &= 2.78516389 \times 10^{-1}, \\
		a_{2}^* &= -1.17431390 \times 10^{-2}, & a_{3}^* &= -4.31112393 \times 10^{-1}, \\
		a_{4}^* &= 3.83571691 \times 10^{-2}, & a_{5}^* &= 3.89152895 \times 10^{-1}, \\
		a_{6}^* &= 5.01744506 \times 10^{-2}.
	\end{aligned}
\end{equation}

These critical points correspond to equilibrium states within the Rastall gravity framework, providing insights into the universe's dynamical behavior. Below, we analyze their physical implications:

\begin{itemize}
	\item \textbf{Matter-Dominated Epoch and Density Perturbations:} The parameters \( x_1^* \) through \( x_7^* \) and \( x_9^* \) are associated with the evolution of matter density perturbations. The positive values, such as \( x_1^* = 2.82692378 \times 10^{-1} \) and \( x_6^* = 6.33124322 \times 10^{-2} \), suggest stable perturbations during the matter-dominated phase, potentially corresponding to structure formation. Conversely, negative values, like \( x_3^* = -3.88706153 \times 10^{-2} \) and \( x_9^* = -6.91572548 \times 10^{-3} \), indicate decaying modes, which may signify collapsing overdense regions.
	
	\item \textbf{Dark Energy and Cosmic Acceleration:} The values \( x_5^* = -3.92961493 \times 10^{-1} \) and \( x_{10}^* = 2.86583679 \) hint at contributions from dark energy or modified gravity effects. The large magnitude of \( x_{10}^* \) suggests a significant influence of dark energy within the Rastall framework, potentially leading to an accelerated expansion phase.
	
	\item \textbf{Anisotropic Stress and Isocurvature Perturbations:} The parameters \( x_8^* = -2.73559806 \times 10^{3} \) and \( x_{13}^* = 1.11530821 \) may be associated with anisotropic stress or isocurvature perturbations. The large magnitude of \( x_8^* \) suggests a pronounced impact on cosmic structure formation, which could be tied to the unique characteristics of Rastall gravity.
	
	\item \textbf{Metric Potentials and Curvature Perturbations:} The coefficients \( a_1^* \) through \( a_6^* \) govern the evolution of metric perturbations. The positive values, such as \( a_1^* = 2.78516389 \times 10^{-1} \) and \( a_5^* = 3.89152895 \times 10^{-1} \), indicate growing modes, which enhance gravitational potential wells, facilitating structure growth. Conversely, negative values like \( a_2^* = -1.17431390 \times 10^{-2} \) suggest decaying modes, which may contribute to smoothing inhomogeneities in the universe.
	
	\item \textbf{Dynamical Stability Analysis:} To assess whether these critical points represent stable attractors, saddle points, or repellers, we must evaluate the eigenvalues of the Jacobian matrix associated with the dynamical system. A negative real part in all eigenvalues would confirm stability, implying that the universe could settle into these states over time. Conversely, positive eigenvalues would indicate instability, suggesting evolutionary transitions or deviations from equilibrium.
\end{itemize}

Further numerical simulations and perturbative analyses are essential to validate these interpretations and to explore the implications of Rastall gravity on cosmological evolution in greater detail.

These critical points represent the equilibrium configurations of the dynamical system associated with perturbed Rastall gravity. Each critical point corresponds to a specific set of values for the perturbation variables and parameters, offering insights into different potential steady states or phases of the system.

The process of solving the differential equations and computing the Jacobian matrix highlights the complexity and depth of the model. The analysis of these critical points will provide valuable information regarding the stability and physical interpretation of the solutions within the framework of perturbed Rastall gravity. Future work will focus on evaluating the stability of these points and their implications for the theoretical and observational aspects of cosmology.

\section{Numerical Analysis}

To evaluate the success of the model under study, we perform a series of Markov-chain Monte Carlo (MCMC) runs, using the public code { MontePython-v3}\footnote{\url{https://github.com/brinckmann/montepython_public}}\cite{54,55}, which we interface with our modified version of { CLASS}~\cite{56,57}. To test the convergence of the MCMC chains, we use the Gelman-Rubin \cite{58} criterion $|R -1|\!\lesssim\!0.01$. To post-process the chains and plot figures we use {\sf GetDist} \cite{59}. 
All observational data where used in this paper are:\\
$\bullet$ {CMB data}:
We used the latest large-scale cosmic microwave background (CMB) temperature and
polarization angular power spectra from the final release of Planck 2018 plikTTTEEE+lowl+lowE
\cite{43}. \\
$\bullet$ {Lensing}: we consider the 2018 CMB lensing reconstruction power spectrum data,
obtained with a CMB trispectrum analysis in \cite{Aghanim1}.

$\bullet$ {BAO data}:
We also used the various measurements of the Baryon Acoustic Oscillations (BAO) from:\cite{Ratra1}
\begin{table}
	\centering
	\caption{12 BAO data.}\label{tab:bao}
	\scriptsize
	\begin{tabular}{lccc}
		\hline
		$z$ & Measurement{a} & Value \\
		\hline
		\hline
		$0.122$ & $D_V\left(r_{s,{\rm fid}}/r_s\right)$ & $539\pm17$ \\
		$0.38$ & $D_M/r_s$ & 10.23406 \\
		$0.38$ & $D_H/r_s$ & 24.98058 \\
		$0.51$ & $D_M/r_s$ & 13.36595 \\
		$0.51$ & $D_H/r_s$ & 22.31656 \\
		$0.698$ & $D_M/r_s$ & 17.85823691865007 \\
		$0.698$ & $D_H/r_s$ & 19.32575373059217 \\
		$0.835$ & $D_M/r_s$ & $18.92\pm0.51$ \\
		$1.48$ & $D_M/r_s$ & 30.6876 \\
		$1.48$ & $D_H/r_s$ & 13.2609 \\
		$2.334$ & $D_M/r_s$ & 37.5 \\
		$2.334$ & $D_H/r_s$ & 8.99 \\
		\hline
	\end{tabular}
\end{table}
were $D_V$, $r_s$, $r_{s, {\rm fid}}$, $D_M$, $D_H$, and $D_A$ have units of Mpc.

$\bullet$ The 12 BAO measurements listed in Table \ref{tab:bao} cover the redshift range $0.122 \leq z \leq 2.334$. The quantities listed in Table 1 are described as follows:

$\bullet$ $D_V(z)$: Spherically averaged BAO distance, $D_V(z)=[czH(z)^{-1}D^2_M(z)]^{1/3}$, where $c$ is the speed of light and the angular diameter distance $D_A(z) = D_M(z)/(1+z)$ with $D_M(z)$ defined in the following\\
$\bullet$ $D_H(z)$: Hubble distance, $D_H(z)=c/H(z)$\\
$\bullet$ $r_s$: Sound horizon at the drag epoch, $r_{s, {\rm fid}}=147.5$ Mpc.\\
$\bullet$ $D_M(z)$: Transverse comoving distance,
\begin{equation}
	\label{eq:DM}
	D_M(z) = 	D_C(z) \ \ \ \  \text{if}\ \Omega_{k0} = 0,\\
\end{equation}
where the comoving distance
\begin{equation}
	\label{eq:gz}
	D_C(z) = c\int^z_0 \frac{dz'}{H(z')}.
\end{equation}
\\
$\bullet$ {CC data}: The cosmic chronometer (CC) data covering the redshift $0.07 < z < 1.965$.(Table 2)\\

\begin{table}[t!]
	\tiny
	\centering
	\caption{32 $H(z)$ (CC) data.}\label{tab:hz}
	\begin{tabular}{lcc}
		\hline
		$z$ & $H(z)$ & $\sigma$\\
		\hline
		0.07 & $69.0$ & 19.6\\
		0.09 & $69.0$ & 12.0\\
		0.12 & $68.6$ & 26.2\\
		0.17 & $83.0$ & 8.0\\
		0.2 & $72.9$ & 29.6\\
		0.27 & $77.0$ & 14.0\\
		0.28 & $88.8$ & 36.6\\
		0.4 & $95.0$ & 17.0\\
		0.47 & $89.0$ & 50.0\\
		0.48 & $97.0$ & 62.0\\
		0.75 & $98.8$ & 33.6\\
		0.88 & $90.0$ & 40.0\\
		0.9 & $117.0$ & 23.0\\
		1.3 & $168.0$ & 17.0\\
		1.43 & $177.0$ & 18.0\\
		1.53 & $140.0$ & 14.0\\
		1.75 & $202.0$ & 40.0\\
		0.1791 & 74.91 & 4.00\\
		0.1993 & 74.96 & 5.00\\
		0.3519 & 82.78 & 14\\
		0.3802 & 83.0 &  13.5\\
		0.4004 & 76.97 &  10.2\\
		0.4247 & 87.08 &  11.2\\
		0.4497 & 92.78 &  12.9\\
		0.4783 & 80.91 &  9\\
		0.5929 & 103.8 & 13\\
		0.6797 & 91.6 & 8\\
		0.7812 & 104.5 & 12\\
		0.8754 & 125.1 & 17\\
		1.037 & 153.7 & 20\\
		1.363 & 160.0 & 33.6\\
		1.965 & 186.5 & 50.4\\
		\hline
	\end{tabular}
\end{table}
$\bullet$ Pantheon catalog:
We used updated the Pantheon + Analysis catalog consisting of 1701 SNe Ia covering the redshift range $0.001 < z < 2.3$\cite{42}.\\
Table 3 shows the  flat priors for the cosmological parameters based on (\cite{Aghanim}; \cite{Aghanim1})
\begin{table}
	\begin{center}
		\caption{Flat priors for the cosmological parameters in GRG coupled with neutrinos.}
		\resizebox{0.2\textwidth}{!}{
			\begin{tabular}{cc}
				\hline 
				Parameter                    & Prior\\
				\hline 
				$\tau$                       & $[0,0.8]$\\
				$n_s$                        & $[0.8,1.2]$\\
				$\log[10^{10}A_{s}]$         & $[1.6,3.9]$\\
				$100\theta_{MC}$             & $[0.5,10.5]$\\ 
				$\Omega_{\nu}$               & $[0,0.1]$\\
				$\alpha$               & $[-1,1]$\\
				\hline 
			\end{tabular}
		}
	\end{center}
	\label{tab:priors}
\end{table}
Also, we used the Akaike Information Criteria (AIC).
The Akaike Information Criterion (AIC) is a statistical measure used to compare different statistical models based on their ability to fit the data while balancing the complexity of the model. The  AIC equation is:

\begin{equation}\label{key}
	AIC = \chi _{\min }^2 + 2\gamma	
\end{equation}
In these equations$\chi _{\min }^2  $ is the minimum value of $ {\chi ^2} $, $\gamma$ is the number of parameters of the given model.

\section{The Early Universe}
In the study of the early universe, alternative gravitational theories like Rastall gravity offer unique perspectives beyond standard general relativity. Rastall gravity introduces non-conservative modifications to the gravitational field equations, dynamically influenced by the universe's matter content. In the early universe, Rastall gravity may deviate from standard cosmological scenarios, impacting the evolution of density perturbations and overall structure formation. The modified gravitational action could provide insights into cosmological puzzles such as dark energy. Analyzing Rastall gravity's implications involves examining its effects on key cosmological parameters, including primordial nucleosynthesis, cosmic microwave background radiation, and large-scale structure formation. In early universe cosmology, Rastall gravity introduces distinctive characteristics, altering the Integrated Sachs-Wolfe effect and the Cosmic Microwave Background spectrum. The non-conservative nature of Rastall gravity leads to unique signatures in the CMB temperature anisotropy, deviating from expectations based on general relativity. This departure manifests as distinctive temperature fluctuations in the CMB spectrum.
During the hot and dense conditions of the early universe, neutrinos are produced in copious amounts and are kept in thermal equilibrium with other particles. As the universe expands and cools, the temperature decreases, and the neutrinos undergo a phase transition known as the neutrino decoupling. This transition marks the point at which neutrinos become non-relativistic.

In the framework of generalized Rastall gravity, where the energy-momentum tensor's non-conservativity persists, the gravitational coupling dynamically responds to both the matter content of the universe and the presence of neutrinos with non-negligible masses. This introduces a nuanced perspective on the cosmic interplay between gravitational forces and neutrino contributions.
\subsection{Coupled neutrinos with Rastall gravity}
The non-conservative nature of Rastall gravity, coupled with the inclusion of massive neutrinos, becomes particularly relevant in the early universe. Deviations from standard cosmological scenarios, influenced by the interplay between modified gravity and massive neutrinos, can impact the evolution of density perturbations, alter the dynamics of structure formation, and potentially offer solutions to cosmological enigmas.
In this section, we first investigate the effect of coupling non - relativistic neutrinos to Rastall gravity model in CMB power spectrum. After that, by using the CMB + Lensing data and use the MCMC method, we estimate the $\Omega_{\rm m}$, (where $\Omega_{\rm m} = \Omega_{\rm b}+\Omega_{\rm cdm}+\Omega_{\nu}$) and $H_{0}$. Moreover, we investigate the perturbed Rastall gravity on alleviating the $H_{0}$ tension. In the early universe, where quantum effects and high-energy phenomena play a pivotal role, Rastall gravity introduces alterations to the standard gravitational equations. By allowing for the non-conservation of energy-momentum, this modified gravity framework offers a novel perspective on the evolution of the cosmos.
The Sachs-Wolfe effect originates from the interaction between gravitational potentials and photons as they traverse through evolving gravitational fields in an expanding universe.The study of coupled neutrinos with the Rastall gravity and perturbed Rastall gravity reveals that gravitational fields undergo changes that leave an imprint on the CMB power spectrum. Figure 1 illustrates the Sachs-Wolfe effect for $l=10$ to $l=100$, where the comparison between coupled  Rastall gravity with neutrinos and perturbed Rastall gravity  and the $\Lambda$CDM model is presented.
If the coupling between dark energy and other cosmic components persists into late cosmic times, the LISW effect may be affected. Changes in the late-time acceleration of the universe due to coupled Rastall gravity could leave imprints on the CMB power spectrum at smaller angular scales. The presence of coupled Rastall gravity might influence the behavior of acoustic oscillations in the primordial plasma. This can lead to alterations in the positions and amplitudes of the peaks and troughs in the CMB power spectrum associated with acoustic oscillations. Figure 1 provides a compelling visual exploration of the Cosmic Microwave Background (CMB) power spectrum, unveiling distinct features in coupled Rastall gravity with neutrinos, and the $\Lambda $CDM model.

The plotted curves elegantly illustrate the variations in the power spectrum, offering a nuanced perspective on the impact of different gravitational frameworks on the observed CMB signatures. Specifically, the inclusion of neutrinos in coupled Rastall gravity introduces subtle shifts in the peaks, indicative of the intricate interplay between gravitational interactions and neutrino contributions.

As we can see in Fig 1, for $l \ge 200$, the first peak shifts down with respect to the $\Lambda $CDM model. We can conclude if we consider the models which are Rastall gravity coupled with neutrinos create a change in the distribution of matter in the universe.
\begin{figure}
	\includegraphics[width=14 cm]{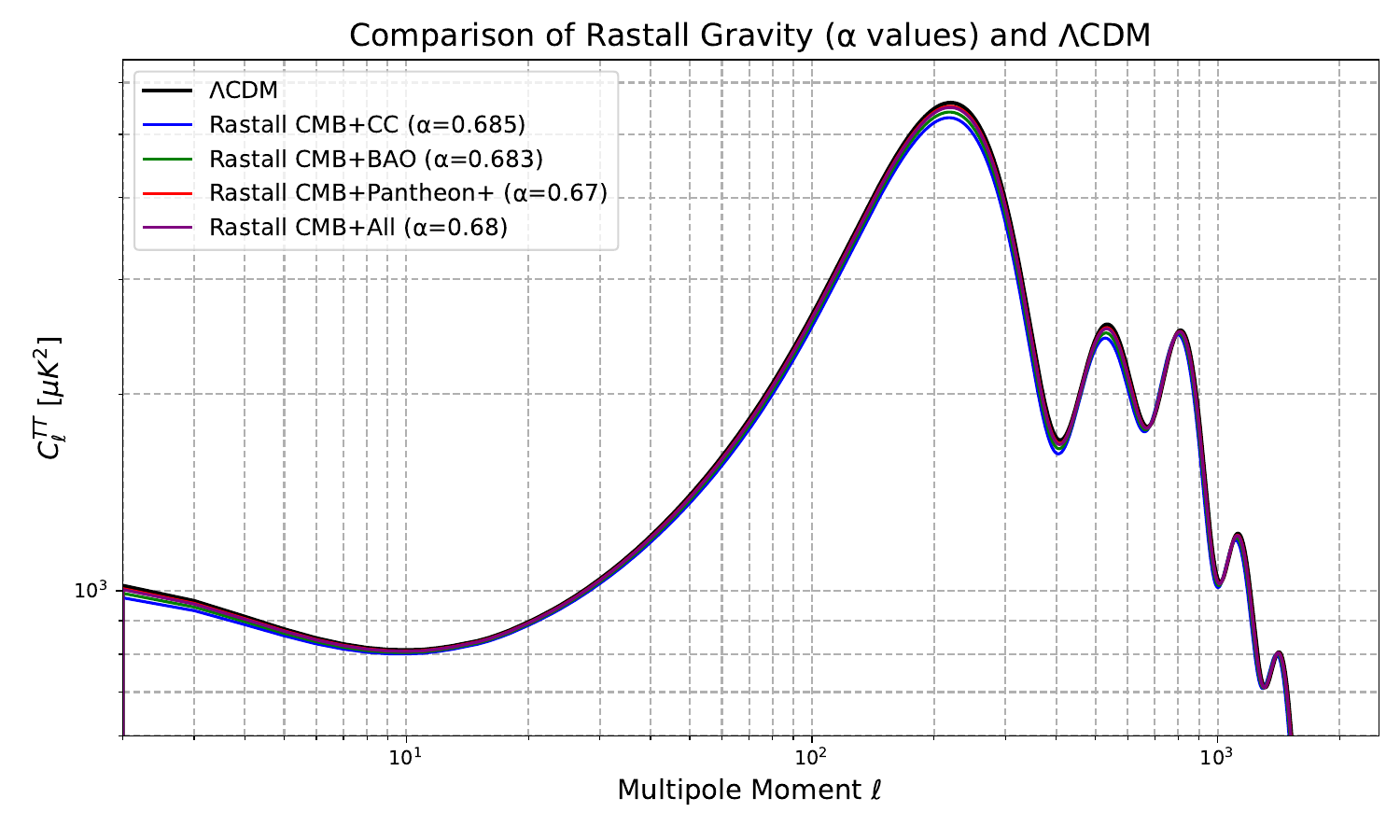}
	\vspace{-0.12cm}
	\caption{\small{Comparison the CMB power spectrum  between coupled perturbed Rastall gravity with neutrinos and $\Lambda $CDM model. }}\label{fig:omegam2}
\end{figure}
We can  put constraints on the following cosmological parameters: the Baryon energy density ${\Omega _b}{h^2}$, the cold dark matter energy density $\Omega_{c}h^{2}$, the neutrino density ${\Omega _\nu}$, the ratio of the sound horizon at decoupling to the angular diameter distance to last scattering $\theta_{MC}$, the optical depth to reionization $\tau$, the amplitude and the spectral index of the primordial scalar perturbations $A_{s}$ and $n_{s}$. Table 4 presents these cosmological parameters. These values provide crucial insights into the dynamics and evolution of the early universe under the framework of coupled perturbed Rastall gravity. Figure 2 provides a comparison of the obtained values for key cosmological parameters in the early universe.

An increase in the coupling strength parameter \(\alpha\), representing the interaction between neutrinos and dark energy (or equivalently the non-conservation of the energy-momentum tensor in Rastall gravity), leads to a noticeable suppression of the first acoustic peak in the CMB temperature power spectrum. Physically, this effect arises from modifications in the gravitational potentials due to the interaction, which alters the evolution of perturbations prior to recombination. The coupling reduces the effective driving force for acoustic oscillations in the photon-baryon fluid, leading to a smaller amplitude of the first peak. This is analogous to the effect of increasing the effective neutrino mass, which also suppresses the acoustic peaks. Additionally, the altered dynamics of gravitational potentials can affect the Integrated Sachs-Wolfe (ISW) effect, further contributing to the reduction in peak height. Numerically, we observe that for \(\alpha = 0.4\), the amplitude of the first peak is lower than in the standard \(\Lambda\)CDM model, confirming this theoretical expectation.

\begin{table*}
	\caption{Results obtain for coupled perturbed Rastall gravity with neutrinos for $H_{0}$, $\Omega_{\rm m}$, $\Omega_b h^2$, $\Omega_c h^2$, $n_s$, ${\rm{ln}}(10^{10} A_s)$,$\tau$ and $100\theta_{MC}$ obtained values for the  Early universe.}
	\begin{tabular}{l c c c}
		
		\hline
		Parameter & Lensing & CMB & CMB + Lensing\\
		\hline
		{\boldmath$\Omega_b h^2$} & $0.02194^{+0.00040}_{-0.00035}$ & $0.02225\pm 0.00024$ & $0.02237\pm 0.00018$\\
		
		{\boldmath$\Omega_c h^2$} & $0.1199\pm 0.0040$ & $0.1200\pm 0.0037$ & $0.1187\pm 0.0025$\\
		
		{\boldmath$100\theta_{MC}$} & $1.04075\pm 0.00060$ & $1.04089\pm 0.00057$ & $1.04108\pm 0.00039$\\
		
		{\boldmath$\tau$} & $0.0510\pm 0.0081$ & $0.0553^{+0.0054}_{-0.0080}$ & $0.0561^{+0.0056}_{-0.0082}$\\
		
		{\boldmath${\rm{ln}}(10^{10} A_s)$} & $3.035\pm 0.021$ & $3.046^{+0.015}_{-0.018}$ & $3.045^{+0.014}_{-0.018}$\\
		
		{\boldmath$n_s$} & $0.956^{+0.016}_{-0.014}$ & $0.9679\pm 0.0087$ & $0.9651\pm 0.0064$\\
		
		$H_0$ & $68.6\pm 3.3$ & $69.3\pm 1.8$ & $69.1\pm 1.5$\\
		
		$\Omega_m$ & $0.348^{+0.017}_{-0.048}$ & $0.3092\pm 0.0084$ & $0.3111\pm 0.0071$\\
		\hline
	\end{tabular}
\end{table*}

\begin{figure*}
	\includegraphics[width=15 cm]{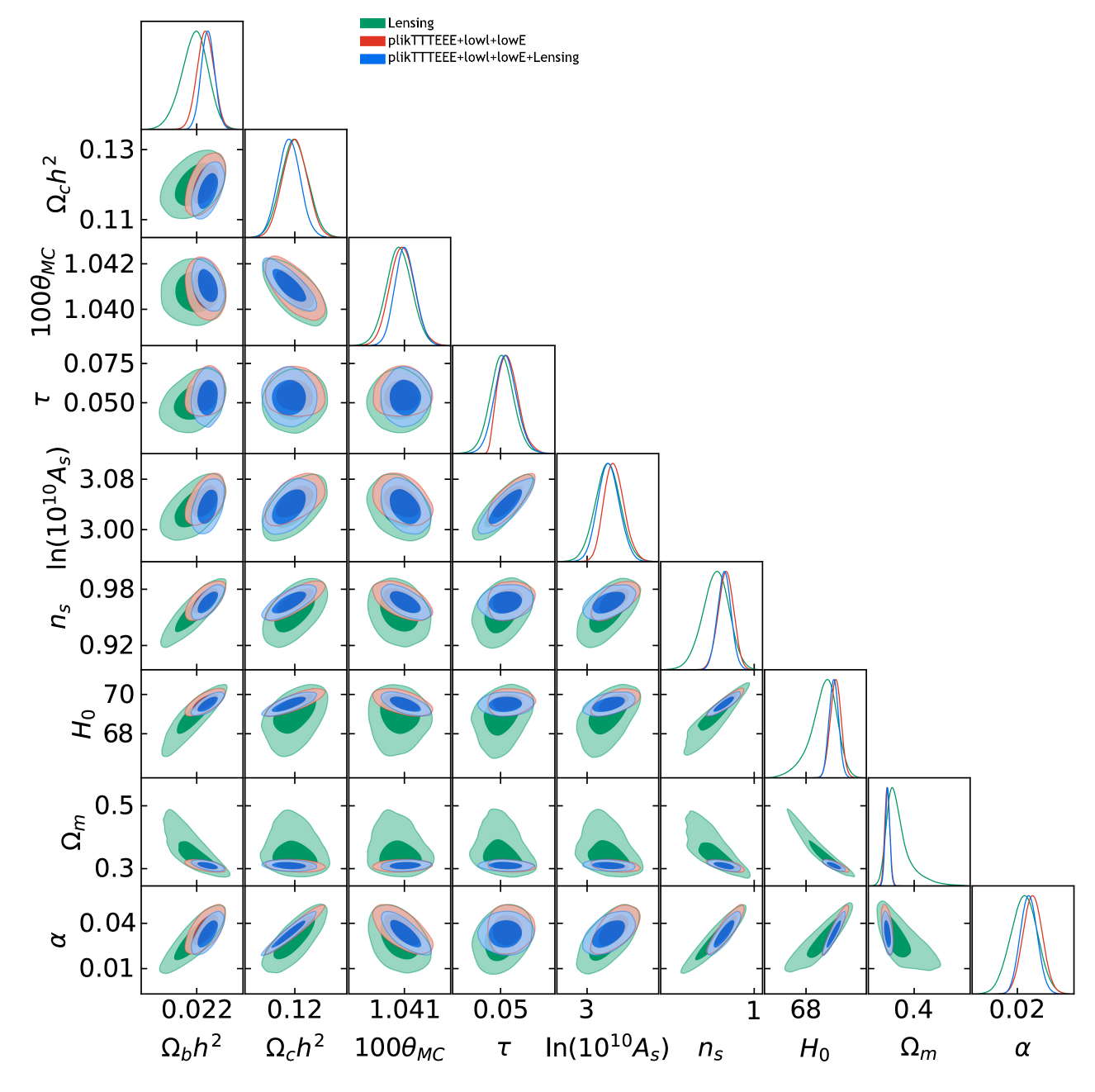}
	\vspace{-0.12cm}
	\caption{\small{Comparison of $H_{0}$, $\Omega_{\rm m}$, $\Omega_b h^2$, $\Omega_c h^2$, $n_s$, ${\rm{ln}}(10^{10} A_s)$,$\tau$ and $100\theta_{MC}$ obtained values for the  Early universe. }}\label{fig:omegam2}
\end{figure*}

\section{Tension Analysis with R22 and Planck 2018}

To assess the tension between our results and those from the Planck 2018 and R22 surveys, we compared the Hubble constant ($H_0$) values obtained from our analyses. The results are summarized as follows:

\begin{table}[h]
	\centering
	\caption{Calculated tensions between our results and values from Planck 2018 and R22.}
	\begin{tabular}{lccc}
		\hline
		& Planck 2018 & R22 \\
		\hline
		Lensing & $0.36\sigma$ & $1.29\sigma$ \\
		CMB & $1.01\sigma$ & $1.76\sigma$ \\
		CMB + Lensing & $1.08\sigma$ & $2.16\sigma$ \\
		\hline
	\end{tabular}
\end{table}

\subsection{Comparison with Planck 2018}

The tension between our $H_0$ values and those from Planck 2018 is relatively modest. Specifically:

\begin{itemize}
	\item For the Lensing data, the tension is approximately $0.36\sigma$.
	\item For the CMB data, the tension is about $1.01\sigma$.
	\item For the combined CMB + Lensing data, the tension is approximately $1.08\sigma$.
\end{itemize}

These results are shown in table 5, indicate that the values are reasonably consistent with Planck 2018, with tensions generally within $1\sigma$ to $1.08\sigma$.

\subsection{Comparison with R22}

In contrast, the tension between our results and R22 is more pronounced:

\begin{itemize}
	\item For the Lensing data, the tension is about $1.29\sigma$.
	\item For the CMB data, the tension is approximately $1.76\sigma$.
	\item For the combined CMB + Lensing data, the tension reaches $2.16\sigma$.
\end{itemize}

These tensions suggest a more significant discrepancy with R22, particularly for the combined dataset.

\noindent This analysis highlights the varying levels of tension between different observational datasets and underscores the importance of considering these differences in cosmological model interpretations.

The contours delineate the 68\% confidence limits, illuminating the regions of parameter space consistent with the observed data. The striking interplay between CMB and Lensing datasets is evident, showcasing the complementary nature of these cosmological probes.

In the realm of CMB, the contours encapsulate the nuanced interplay of temperature and polarization fluctuations, unraveling the intricate imprint of early universe physics. Lensing, on the other hand, provides a unique lens (pun intended) into the gravitational effects on cosmic structures, contributing vital constraints to our cosmological model.

The combined constraints, depicted with contours that gracefully merge the distinctive signatures of CMB and Lensing, yield a refined understanding of cosmological parameters. This synergy enhances the precision of parameter determination, contributing to the advancement of our comprehension of the universe.

Table 6 presents a comparison of the $\chi^2$ values between the $\Lambda$CDM model and perturbed Rastall gravity coupled with neutrinos for various dataset combinations. The datasets considered include the Cosmic Microwave Background (CMB), Lensing, and their combination (CMB+Lensing). This comparison helps in evaluating the fit of each model to the observational data and assessing the performance of perturbed Rastall gravity with neutrinos relative to the standard $\Lambda$CDM cosmology. The table highlights the $\chi^2_{\rm tot}$ values, which reflect the overall goodness of fit, as well as the individual contributions from different components, such as CMB and Lensing, for each model.

\begin{table}
	\tiny
	\caption{{\small $\chi^2_{}$ comparison between $\Lambda$CDM and perturbed Rastall gravity coupled with neutrinos for CMB, Lensing, and CMB+Lensing dataset combinations.}}
	\begin{center}
		\resizebox{0.8\textwidth}{!}{  
			\begin{tabular}{ c |c c c } 
				\hline
				\hline
				$\Lambda$CDM  & CMB & Lensing & CMB+Lensing \\ 
				\hline
				$\chi^2_{\rm  tot}$ & $-$ & $-$ & $2776.250$ \\
				$\chi^2_{\rm  CMB}$ & $2778.132$ & $-$ & $2768.123$ \\
				$\chi^2_{\rm lensing}$ & $-$ & $8.254$ & $8.127$ \\
				\hline
				\hline
		perturbed	Rastall gravity coupled with neutrinos  & CMB & Lensing & CMB+Lensing \\ 
				\hline
				$\chi^2_{\rm  tot}$ & $-$ & $-$ & $2770.62$ \\
				$\chi^2_{\rm  CMB}$ & $2773.134$ & $-$ & $2762.775$ \\
				$\chi^2_{\rm  lensing}$ & $-$ & $7.845$ & $7.845$ \\
				\hline
				\hline
			\end{tabular}
		}
	\end{center}
	\label{table_chi_cmb_lensing}
\end{table}

\section{The Local universe}

In the Local universe, the behavior of neutrinos, which are nearly massless particles, becomes pivotal in the cosmic narrative (\cite{Lesgourgues}; \cite{lesgourgues2006}). When considered within the framework of Rastall gravity, a theory proposing a departure from the conventional conservation laws for the energy-momentum tensor—neutrinos exhibit distinctive effects, significantly shaping the late-time evolution of the cosmos.

As the universe expands, neutrinos transition from a relativistic to a non-relativistic state, impacting their role in cosmic structure formation. In the context of Rastall gravity, where the energy-momentum tensor is not strictly conserved, neutrinos experience additional gravitational forces that alter their trajectories and clustering patterns. This departure from standard gravitational predictions has observable consequences on large-scale structures, influencing phenomena like cosmic microwave background anisotropies and galaxy clustering.
The study of neutrinos within this modified gravitational framework provides a means to probe the fundamental principles underlying cosmic acceleration. The coupling between neutrinos and Rastall gravity may influence the late-time dynamics, affecting the cosmic web, galaxy clustering, and other large-scale structures. These modifications, when considered in the context of the Hubble tension, provide an avenue for reconciling discrepancies between local measurements and those inferred from early universe observations.

Another crucial approach for estimating the Hubble parameter ($H_0$) involves the use of Cosmic Chronometers (CC) and Baryon Acoustic Oscillation (BAO) data. These local measurements offer valuable insights into the expansion history of the universe; however, a persistent discrepancy exists when comparing them with values derived from early universe observations, such as those from the Cosmic Microwave Background (CMB). This discrepancy, widely known as the *Hubble tension*, indicates a potential mismatch or systematic difference between local and early-universe determinations of $H_0$.

\vspace{0.3cm}
\noindent
$\bullet$ \textbf{CMB + BAO:} \\
The combination of CMB and BAO data yields $H_0 = 70.25 \pm 2.1$ km/s/Mpc at 68\% confidence level. This result shows a mild tension of $1.33\,\sigma$ with the Planck 2018 value ($67.4 \pm 0.5$ km/s/Mpc) and $1.13\,\sigma$ with the R22 result ($73.04 \pm 1.04$ km/s/Mpc).

\vspace{0.3cm}
\noindent
$\bullet$ \textbf{CMB + CC:} \\
The combination of CMB and CC data gives $H_0 = 70.86 \pm 2.26$ km/s/Mpc. The derived Hubble constant shows a $1.49\,\sigma$ tension with Planck 2018 and a $0.87\,\sigma$ tension with R22, indicating relatively good consistency with both measurements.

\vspace{0.3cm}
\noindent
$\bullet$ \textbf{CMB + Pantheon+:} \\
This combination leads to $H_0 = 71.09 \pm 2.3$ km/s/Mpc, exhibiting a $1.57\,\sigma$ tension with Planck 2018 and a $0.78\,\sigma$ tension with R22.

\vspace{0.3cm}
\noindent
$\bullet$ \textbf{CMB + All datasets:} \\
When all datasets are combined, the result is $H_0 = 70.23 \pm 2.01$ km/s/Mpc, with a tension of $1.34\,\sigma$ relative to Planck 2018 and $1.16\,\sigma$ with respect to R22.

\vspace{0.4cm}
\noindent
These results are summarized in Tables~\ref{tab:combined_results} and \ref{tab:H0_tension}. Table~\ref{tab:combined_results} lists the mean values and 1$\sigma$ uncertainties for the key cosmological parameters derived from different combinations of observational datasets. Table~\ref{tab:H0_tension} quantifies the Hubble tension by comparing the inferred $H_0$ values with the standard Planck and R22 benchmarks, expressed in units of standard deviation.

\vspace{0.4cm}
\noindent
Figure 3 illustrate the variation in the obtained values of $H_0$, $\Omega_m$, $\Omega_b h^2$, and $\Omega_c h^2$ within the local universe context. Table VII specifically addresses the impact of coupling neutrinos with perturbed Rastall gravity on the $H_0$ value. Furthermore, Table VIII provides a comprehensive comparison of key $H_0$ tension in local universe modeling scenarios. These results are in broad agreement with \cite{Y1, Y2, Y3, Y4}

\begin{table*}[h]
	\centering
	\renewcommand{\arraystretch}{1.2}
	\begin{tabular}{l c c c c c}
		\hline
		Parameter &  CMB + CC & CMB +  BAO & CMB +  Pantheon+ & CMB + All \\
		\hline
		$\Omega_b h^2$ &  $0.02224\pm 0.00024$ & $0.02221\pm 0.00022$ & $0.02234\pm 0.00019$ & $0.02236\pm 0.00018$ \\
		$\Omega_c h^2$ &  $0.1200\pm 0.0037$ & $0.1190\pm 0.0028$ & $0.1184\pm 0.0034$ & $0.1180\pm 0.0029$ \\
		$100\theta_{MC}$ &  $1.04089\pm 0.00057$ & $1.04099\pm 0.00050$ & $1.04103^{+0.00044}_{-0.00051}$ & $1.04116\pm 0.00042$ \\
		$\tau$ &  $0.0544\pm 0.0076$ & $0.0552^{+0.0054}_{-0.0078}$ & $0.0548^{+0.0050}_{-0.0079}$ & $0.0551\pm 0.0074$ \\
		$\ln(10^{10} A_s)$ &  $3.044\pm 0.018$ & $3.043^{+0.014}_{-0.017}$ & $3.039^{+0.015}_{-0.018}$ & $3.041\pm 0.016$ \\
		$n_s$ &  $0.9677\pm 0.0087$ & $0.9657\pm 0.0071$ & $0.9669\pm 0.0076$ & $0.9637\pm 0.0070$ \\
		$H_0$ &  $70.86\pm 2.26$ & $70.25\pm 2.1$ & $71.09\pm2.3$ & $70.23\pm 2.01$ \\
		$\Omega_m$ &  $0.314\pm 0.008$ & $0.306\pm 0.0081$ & $0.305\pm 0.007$ & $0.303\pm 0.0069$ \\
			$\alpha$ &  $0.685\pm 0.17$ & $0.683\pm 0.15$ & $0.67\pm 0.24$ & $0.68\pm 0.12$ \\
		\hline
	\end{tabular}
	\caption{Parameter constraints at  68\% confidence level  for different dataset combinations.}
	\label{tab:combined_results}
\end{table*}
\begin{table*}[h]
	\centering
	\renewcommand{\arraystretch}{1.3}
	\begin{tabular}{l c c}
		\hline
		Dataset & Tension with Planck 2018 ($\sigma$) & Tension with R22 ($\sigma$) \\
		\hline
		CMB + CC         & 1.49 & 0.87 \\
		CMB + BAO        & 1.33 & 1.13 \\
		CMB + Pantheon+  & 1.57 & 0.78 \\
		CMB + All        & 1.34 & 1.16 \\
		\hline
	\end{tabular}
	\caption{Hubble tension between derived $H_0$ values and Planck 2018 ($67.4 \pm 0.5$ km/s/Mpc) and R22 ($73.04 \pm 1.04$ km/s/Mpc), reported in units of standard deviation.}
	\label{tab:H0_tension}
\end{table*}

\begin{figure*}
	\includegraphics[width=15 cm]{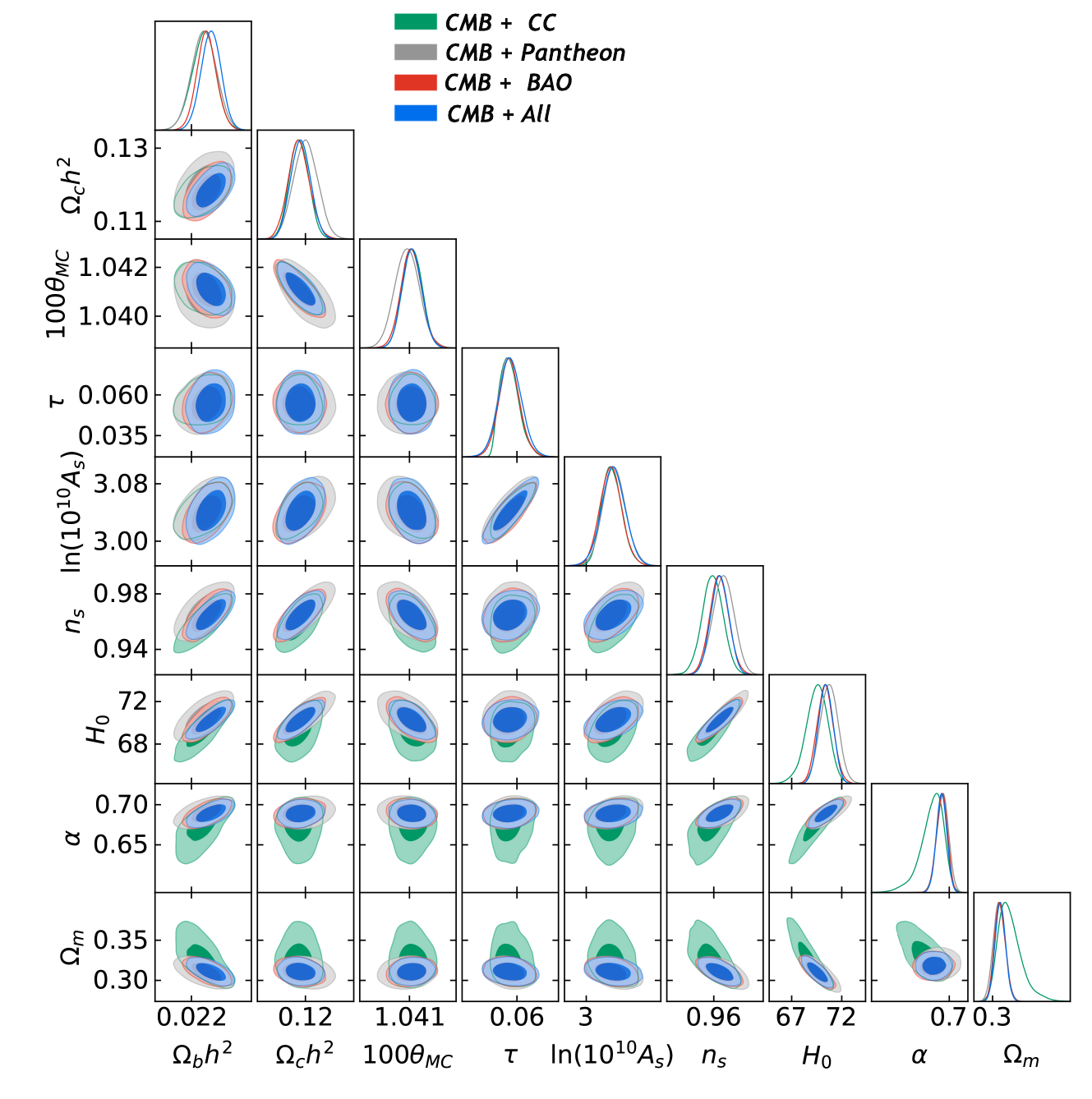}
	\vspace{-0.12cm}
	\caption{\small{Comparison of $\Omega_{\rm b}h^{2}$, $\Omega_{\rm c}h^{2}$, $H_{0}$, $\Omega_{\rm m}$ obtained values for the  Local universe. }}\label{fig:omegam2}
\end{figure*}
Table 9 presents a comparison of the $\chi^2$ values between the $\Lambda$CDM model and perturbed Rastall gravity coupled with neutrinos for various dataset combinations. The datasets included in this comparison are  (CC), (BAO), and their combined dataset (CC+BAO). This table evaluates the goodness of fit for each model with respect to these datasets, providing insights into how well  perturbed Rastall gravity with neutrinos performs relative to the standard $\Lambda$CDM cosmology across different observational data sets.

\begin{table*}
	\caption{{\small $\chi^2_{}$ comparison between $\Lambda$CDM and Pertubed Rastall gravity model for the different dataset combinations explored in this work. CMB+all refers to Planck+BAO+CC+Pantheon+Lensing.}}
	\begin{center}
		\resizebox{0.85\textwidth}{!}{  
			\begin{tabular}{| c |c| c| c| c |c| } 
				\hline
				\hline
				$\Lambda$CDM  & CMB+CC & CMB+BAO & CMB+Pantheon+ & CMB+all \\ 
				\hline
				$\chi^2_{\rm  tot}$ &  $2795.459$ & $2779.182$ & $3592.124$ & $3639.36$  \\
				$\chi^2_{\rm  CMB}$ & $2768.743$ & $2772.019$ & $2767.712$ & $2779.886$  \\
				$\chi^2_{\rm  CC}$  & $26.716$   & $-$        & $-$        & $27.941$  \\
				$\chi^2_{\rm  BAO}$ & $-$        & $7.163$    & $-$        & $7.577$  \\
				$\chi^2_{\rm  Pantheon+}$& $-$   & $-$        & $824.412$  & $823.956$  \\
				\hline
				\hline
			Pertubed Rastall gravity model  &  CMB+CC & CMB+BAO & CMB+Pantheon+ & CMB+all \\ 
				\hline
				$\chi^2_{\rm  tot}$ & $2784.629$ & $2771.925$ & $3695.631$ & $3606.33$  \\
				$\chi^2_{\rm  CMB}$ &  $2763.531$ & $2766.613$ & $2765.121$ & $2771.029$  \\
				$\chi^2_{\rm  CC}$  &  $21.098$   & $-$        & $-$        & $20.885$  \\
				$\chi^2_{\rm  BAO}$ &  $-$        & $5.312$    & $-$        & $5.068$  \\
		  $\chi^2_{\rm  Pantheon+}$ & $-$         & $-$        & $809.631$  & $809.348$  \\
				\hline
				\hline
			\end{tabular}
		}
	\end{center}
	\label{table_chi}
\end{table*}
We obtained the value of $\alpha = 0.68 \pm 0.12$, which is broadly consistent with the findings of \cite{Lin}, \cite{Yarahmadi}. 

Tables X  present the mean values of the free parameters with 1$\sigma$ error bars for different cosmological models, derived using combined datasets for both the early and the local universe.

In Table X, which focuses on the early universe, the Akaike Information Criterion (AIC) values for the $\Lambda$CDM model and an alternative model (denoted as "The Early Universe") are compared. The $\Lambda$CDM model yields an AIC value of 2782.25, while the alternative model has a lower AIC value of 2778.62. This reduction in AIC indicates that the alternative model offers a better fit to the data, suggesting it may provide a more accurate description of the early universe when considering the trade-off between goodness of fit and model complexity.

Table X presents the corresponding results for the local universe. The AIC for the $\Lambda$CDM model is 3645.36, whereas the alternative model (referred to as "The Local Universe") achieves a notably lower AIC of 3616.33. Once again, the lower AIC supports the conclusion that the alternative model provides a better fit to the data, reinforcing its potential to describe the local universe more effectively than the standard $\Lambda$CDM framework.

Overall, the comparison of AIC values across both datasets suggests that the alternative models outperform the $\Lambda$CDM model in fitting the observational data, highlighting their potential advantage in describing both the early and local universe.

\begin{table}[h!]
	\tiny
	\caption{{\small Mean values of free parameters with 1$\sigma$ error bars for the combined datasets in both the early and local universe scenarios.}}
	\label{table_chi_combined}
	\begin{center}
		\resizebox{1\textwidth}{!}{  
			\begin{tabular}{c|c|ccccccc}
				\hline\hline
				Universe & Model & $\Omega_{\rm de}$ & $\Omega_{\rm m}$ & $\Omega_{\Lambda}$ & $\alpha$ & $\Omega_{\nu}$ & $H_0$ (km/s/Mpc) & AIC \\
				\hline\hline
				\multirow{2}{*}{Early} 
				& $\Lambda$CDM & $-$ & $0.306\pm0.0119$ & $0.685\pm0.028$ & $-$ & $-$ & $68.5\pm1.6$ & $2782.25$ \\
				& Modified (Rastall) & $0.675\pm0.057$ & $0.311\pm0.0071$ & $-$ & $0.035\pm0.007$ & $0.0016$ & $69.1\pm1.5$ & $2778.62$ \\
				\hline
				\multirow{2}{*}{Local} 
				& $\Lambda$CDM & $-$ & $0.326\pm0.032$ & $0.689\pm0.031$ & $-$ & $-$ & $68.7\pm2.6$ & $3645.36$ \\
				& Modified (Rastall) & $0.681\pm0.052$ & $0.309\pm0.0083$ & $-$ & $0.68\pm0.12$ & $0.00312$ & $70.23\pm2.01$ & $3616.33$ \\
				\hline\hline
			\end{tabular}
		}
	\end{center}
\end{table}

	\section*{Physics-Informed Neural Network (PINN) approach}

\subsection{Motivation for Using the Physics-Informed Neural Network (PINN) Model through the Rastall gravity Model to Alleviate the Hubble Tension}

Physics-Informed Neural Network (PINN) provides an innovative tool for the reconstruction of the Hubble parameter \( H(z) \) without relying on traditional numerical methods. The PINN approach is particularly advantageous because it seamlessly integrates known physical constraints, such as the Friedmann equations and the conservation of energy, directly into the training process of the neural network. This not only ensures that the solutions are consistent with physical laws, but it also improves the efficiency of the model by leveraging structured prior information \cite{Raissi2017a , Raissi2017b, Raissi2017c, Raissi2017d, Owhadi2015, Raissi2019, gal2016dropout}. By using the Rastall gravity model within the PINN framework, we aim to uncover a more accurate representation of the Hubble parameter, while simultaneously testing the hypothesis that dark energy dynamics within this model could reduce the observed discrepancy in \( H_0 \).

The flexibility of the PINN model makes it an ideal candidate for addressing the complex non-linear relationships between the cosmological parameters. Through its ability to incorporate observational data, such as Cosmic Chronometer (CC) measurements, alongside the theoretical constraints derived from the Rastall gravity model, the PINN approach allows for a data-driven exploration of cosmological parameters that directly links the physics of dark energy with the observable universe. By combining the Rastall gravity model with PINNs, we can obtain precise reconstructions of \( H(z) \) and assess the potential of the Rastall gravity model in resolving the Hubble tension, providing crucial insights into the nature of dark energy and the overall cosmological framework.

\section{Methodology: Physics-Informed Neural Network for Rastall Gravity}

In this work, we implement a Physics-Informed Neural Network (PINN) to model the evolution of the Hubble parameter $H(z)$ in the framework of Rastall gravity. This approach enables the reconstruction of cosmological dynamics from data while embedding the theoretical constraints from modified gravity directly into the learning algorithm.

\subsection{Neural Network Architecture}

The core of our model is a fully connected feedforward neural network with 5 hidden layers of progressively decreasing size. Specifically, the architecture consists of layers with [1024, 512, 256, 128, 64] neurons. Each hidden layer uses the hyperbolic tangent (\texttt{tanh}) activation function, providing smooth approximations ideal for continuous cosmological functions. To prevent overfitting and enable uncertainty quantification, a dropout layer with a rate of 10\% is inserted after each hidden layer.

The network is constructed using TensorFlow, and its architecture can be summarized in the following code excerpt:

\begin{verbatim}
	model = tf.keras.Sequential()
	model.add(tf.keras.layers.Input(shape=(1,)))
	model.add(tf.keras.layers.Dense(1024, activation='tanh'))
	model.add(tf.keras.layers.Dropout(0.1))
	model.add(tf.keras.layers.Dense(512, activation='tanh'))
	model.add(tf.keras.layers.Dropout(0.1))
	model.add(tf.keras.layers.Dense(256, activation='tanh'))
	model.add(tf.keras.layers.Dropout(0.1))
	model.add(tf.keras.layers.Dense(128, activation='tanh'))
	model.add(tf.keras.layers.Dropout(0.1))
	model.add(tf.keras.layers.Dense(64, activation='tanh'))
	model.add(tf.keras.layers.Dropout(0.1))
	model.add(tf.keras.layers.Dense(1))
\end{verbatim}
\subsection*{Neural Network Architecture}

The architecture of the Physics-Informed Neural Network (PINN) employed in this study is designed to accurately capture the cosmological dynamics across redshift while preserving interpretability and enabling uncertainty quantification. Its structure is as follows:

\begin{itemize}
	
	\item \textbf{Input Layer:} 
	The network accepts a single scalar input representing the redshift, \( z \in [0, 2] \), which serves as the independent cosmological variable. This redshift range is chosen to align with the domain of observational data from cosmic chronometers and other low-redshift probes.
	
	\item \textbf{Hidden Layers:} 
	The core of the model comprises five fully connected (dense) hidden layers with 1024, 512, 256, 128, and 64 neurons, respectively. Each layer employs the hyperbolic tangent (\texttt{tanh}) activation function, which is well-suited for approximating smooth functions with continuous derivatives—a critical requirement for physics-informed models involving differential operators (e.g., \( \frac{dH}{dz} \)) in the loss function. The decreasing width of layers follows a funnel-shaped topology, encouraging a hierarchical abstraction of cosmological features from large to small scales.
	
	\item \textbf{Dropout Regularization:} 
	To improve generalization and facilitate epistemic uncertainty estimation, Dropout regularization is applied after each hidden layer, with a dropout probability of 0.1 (i.e., 10\% of neurons are randomly deactivated during training). This technique serves a dual purpose: (i) it acts as a stochastic regularizer that mitigates overfitting, particularly in sparsely sampled redshift regions, and (ii) it enables approximate Bayesian inference via Monte Carlo Dropout during inference, allowing predictive uncertainty to be estimated from an ensemble of forward passes.
	
	\item \textbf{Output Heads:} 
	The network branches into multiple independent output heads, each corresponding to a physically meaningful prediction:
	
	\begin{itemize}
		\item \textit{Hubble Parameter Head:} A single linear neuron outputs the Hubble expansion rate \( H(z) \). No activation function is applied, allowing the network to freely learn the functional form of \( H(z) \). Physical constraints such as \( H(z) > 0 \) are softly imposed through penalty terms in the custom loss function.
		
		\item \textit{Cosmological Density Head:} A set of three linear neurons predict the constant fractional energy densities \( \Omega_b \), \( \Omega_c \), and \( \Omega_\nu \), corresponding to baryons, cold dark matter, and neutrinos, respectively. These outputs are assumed to be redshift-independent and are regularized using prior-informed penalty terms to ensure consistency with observational bounds and physical viability.
	\end{itemize}
	
\end{itemize}

\subsection{Physics-Informed Loss Function}

The training of the network is guided by a custom loss function that incorporates both data and physical laws. The theoretical form of $H(z)$ in Rastall gravity is given by:

\begin{equation}
	H(z) = H_0 \sqrt{
		\Omega_b (1 + z)^3 +
		\Omega_{cdm} (1 + z)^3 +
		\Omega_{de} (1 + z)^{3(1 + \omega_{de})} \nonumber \\
		+ \Omega_{\nu} (1 + z)^{3(1 + \omega_{\nu})} +
		\Omega_r (1 + z)^4 }.
\end{equation}

We define the custom loss function in TensorFlow using automatic differentiation as follows:

\begin{verbatim}
	def custom_loss(y_true, y_pred, z):
	with tf.GradientTape() as tape:
	tape.watch(z)
	y_pred = model(z)
	dy_pred = tape.gradient(y_pred, z)
	
	# Compute the theoretical H(z) and dH/dz
	y_true_np = H_true(z.numpy().flatten())
	dy_true_np = np.gradient(y_true_np, z.numpy().flatten())
	
	y_true_tensor = tf.convert_to_tensor(y_true_np.reshape(-1, 1),
	dtype=tf.float32)
	dy_true_tensor = tf.convert_to_tensor(dy_true_np.reshape(-1, 1),
	dtype=tf.float32)
	
	loss_h = tf.reduce_mean(tf.square(y_true_tensor - y_pred))
	loss_dh = tf.reduce_mean(tf.square(dy_true_tensor - dy_pred))
	
	return loss_h + loss_dh
\end{verbatim}

This formulation penalizes discrepancies in both the function and its derivative, encouraging the model to satisfy the modified Friedmann equations throughout training.

\subsection{Training and Optimization}

We train the network using the Adam optimizer with a learning rate of $10^{-3}$ for 5000 epochs. The redshift range is sampled from $z \in [0, 2.5]$, normalized to [0, 1] to stabilize training. The model is compiled and trained as follows:

\begin{verbatim}
	optimizer = tf.keras.optimizers.Adam(learning_rate=0.001)
	for epoch in range(epochs):
	with tf.GradientTape() as tape:
	predictions = model(z_tensor)
	loss = custom_loss(H_tensor, predictions, z_tensor)
	gradients = tape.gradient(loss, model.trainable_variables)
	optimizer.apply_gradients(zip(gradients, model.trainable_variables))
\end{verbatim}

\subsubsection*{Gradient-Based Automatic Differentiation for ODE Loss}

To compute the ODE residuals, we employ TensorFlow's automatic differentiation via the \texttt{tf.GradientTape} API. For each sampled redshift \( z \), the Hubble parameter prediction \( H(z) \) is differentiated with respect to \( z \) using the chain rule within a dynamic computation graph. The core routine used is:

\begin{scriptsize}
	\begin{verbatim}
		with tf.GradientTape() as tape:
		tape.watch(z)
		H_pred = model(z)
		dH_dz = tape.gradient(H_pred, z)
	\end{verbatim}
\end{scriptsize}

This allows the neural network to learn a differentiable approximation of \( H(z) \) that satisfies the physical structure of the Rastall model without requiring the explicit conversion of the Friedmann equation into an integral form. The ODE loss is then constructed by minimizing the squared residual between \( dH/dz \) and the right-hand side \( \mathcal{F}(H, z; \alpha) \) of the theoretical evolution equation.

\subsubsection*{Bayesian Uncertainty Estimation via Dropout Sampling}

To assess the epistemic uncertainty in our reconstructed Hubble expansion rate \( H(z) \), we employ a Bayesian approximation technique known as Monte Carlo (MC) Dropout, as introduced in the seminal work by Gal and Ghahramani \cite{gal2016dropout}. While dropout is traditionally utilized as a regularization mechanism during training, it can also be interpreted as a variational approximation to a deep Gaussian process when kept active at inference time. This interpretation allows for principled uncertainty quantification without the computational burden of full Bayesian inference.

In this framework, the neural network is treated as an ensemble of stochastic forward passes, where the predictive mean and variance are estimated through repeated sampling. Formally, given \( T \) forward passes of the PINN with dropout layers enabled during inference, the predictive distribution of \( H(z) \) is approximated by eqs. 12, 13
where each \( H^{(j)}(z) \) is the output of the PINN at redshift \( z \) for the \( j \)-th stochastic realization.

The implementation is realized in TensorFlow as follows:

\begin{scriptsize}
	\begin{verbatim}
		@tf.function
		def predict_with_uncertainty(f_model, z_input, T=100):
		predictions = [f_model(z_input, training=True) for _ in range(T)]
		mean = tf.reduce_mean(predictions, axis=0)
		stddev = tf.math.reduce_std(predictions, axis=0)
		return mean, stddev
	\end{verbatim}
\end{scriptsize}

Here, the `training=True` flag enforces dropout behavior during inference, thereby injecting randomness into each forward pass. The parameter \( T \) controls the number of stochastic evaluations and is set to \( T=100 \) in our experiments, balancing computational tractability with statistical reliability.

This Bayesian approximation yields an empirical uncertainty band for \( H(z) \), which encapsulates model uncertainty arising from limited training data, architectural stochasticity, and intrinsic degeneracy in cosmological reconstructions. The approach is especially beneficial in regions of sparse Cosmic Chronometer (CC) data—such as higher redshifts—where model extrapolations are inherently more uncertain. The dropout-induced variance thus offers a practical and theoretically grounded method to quantify confidence intervals in our reconstructed cosmological trajectories, complementing the error propagation from observational uncertainties.

\subsection{Advantages of the PINN Approach}

This physics-informed machine learning framework offers multiple advantages:
\begin{itemize}
	\item It embeds cosmological equations as a soft constraint during learning.
	\item It enables learning from data while preserving theoretical structure.
	\item It provides predictive uncertainty via dropout-based sampling.
	\item It is easily extensible to other theories, such as modified gravity.
\end{itemize}
\section{Results and Discussion}

In this section, we present the reconstructed cosmological parameters derived from the Physics-Informed Neural Network (PINN) method applied to the Rastall gravity model, accounting for the contribution of neutrinos. The analysis utilizes Cosmic Chronometer (CC) observational data across different redshift bounds: \( z \leq 0.5 \), \( z \leq 1.0 \), \( z \leq 1.5 \), and \( z \leq 2.0 \). The PINN framework was implemented for different data set sizes, \( N = 50, 100, 150, 200 \), for each redshift range. Table XI summarizes the best-fit values of the Hubble constant \( H_0 \), the neutrino density parameter \( \Omega_{\nu} \), and the Rastall model parameter \( \alpha \), including their corresponding uncertainties.

\vspace{0.2cm}
\textbf{Hubble Constant (\( H_0 \))}: The reconstructed values of \( H_0 \) across all redshift ranges lie within the interval \( 69.38 \leq H_0 \leq 71.82 \) km/s/Mpc. The general trend indicates a slight decrease in \( H_0 \) as the number of data points \( N \) increases, especially for higher redshift ranges. At lower redshifts (e.g., \( z \leq 0.5 \)), \( H_0 \) shows relatively higher values with smaller uncertainties, consistent with local observations. For instance, at \( N = 50 \), we observe \( H_0 = 71.82 \pm 2.05 \) km/s/Mpc, while for \( N = 200 \), \( H_0 \) drops to \( 69.38 \pm 2.25 \) km/s/Mpc.

These results are particularly relevant to the ongoing Hubble tension. The local measurement reported by Riess et al. (R22) gives \( H_0 = 73.04 \pm 1.04 \) km/s/Mpc, while the Planck 2018 CMB analysis under the \(\Lambda\)CDM model yields \( H_0 = 67.4 \pm 0.5 \) km/s/Mpc. Our PINN-reconstructed values lie in between these two extremes and tend to favor a mild tension with both. For example, the result \( H_0 = 71.82 \pm 2.05 \) km/s/Mpc at \( z \leq 0.5 \), \( N = 50 \), is approximately \( 1.5\sigma \) lower than R22 and about \( 1.9\sigma \) higher than Planck 2018, based on combined errors. This suggests that Rastall gravity, when constrained by observational Hubble data using PINNs, may offer a partial resolution or alleviation of the Hubble tension.

\vspace{0.2cm}
\textbf{Neutrino Density (\( \Omega_{\nu} \))}: The values of the neutrino density parameter \( \Omega_{\nu} \) display sensitivity to both the redshift limit and the number of data points. At \( z \leq 1.5 \) and \( N = 100 \), \( \Omega_{\nu} \) reaches its maximum value of \( 0.0152 \pm 0.00078 \), indicating that neutrino effects are more significant at intermediate redshifts with moderate data resolution. However, the values at \( N = 150 \) and \( N = 200 \) are notably smaller, reflecting the PINN's tendency to fit smoother, lower neutrino contributions when provided with more extensive data. This supports the interpretation that constraints on \( \Omega_{\nu} \) are model- and data-sensitive, but remain within expected observational limits.

\vspace{0.2cm}
\textbf{Coupling parameter (\( \alpha \))}: The parameter \( \alpha \), which characterizes Coupling parameter  the Rastall gravity framework to neutrinos, was found to lie in the range \( 0.58 \leq \alpha \leq 0.68 \). Notably, \( \alpha \) tends to decrease slightly with both increasing redshift and increasing dataset size, indicating that larger and deeper data may favor a weaker deviation from standard conservation laws. At higher redshifts (e.g., \( z \leq 2.0 \)), \( \alpha \) falls below 0.60 for \( N = 200 \), whereas at \( z \leq 0.5 \), values around \( \alpha \approx 0.67 \) are observed. 
\vspace{0.2cm}
Overall, the PINN-based reconstructions show consistency and robustness across different redshift bounds and dataset sizes. The smooth behavior of the recovered parameters and their agreement with standard cosmological expectations supports the applicability of the PINN method in testing modified gravity models such as Rastall theory. These results also highlight the utility of including neutrino physics in such reconstructions, as they can affect both the background expansion and the effective gravitational dynamics at late times. All these results are in table XI and figures 4,5.
\begin{table}[ht]
	\centering
	\caption{Reconstructed cosmological parameters obtained from the PINN method applied to the Rastall gravity model with neutrinos, evaluated for Cosmic Chronometers (CC) data up to various redshift limits.} 
	\label{tab:PINN_Rastall}
	\begin{tabular}{|c|c|c|c|c|}
		\hline
		\textbf{Redshift Range} & \textbf{N} & \( H_0 \) [km/s/Mpc] & \( \Omega_{\nu} \) & \( \alpha \) \\
		\hline
		\multirow{4}{*}{\( z \leq 0.5 \)} 
		& 50  & \( 71.82 \pm 2.05 \) & \( 0.0026 \pm 0.0002 \)  & \( 0.67 \pm 0.030 \) \\
		& 100 & \( 70.11 \pm 1.85 \) & \( 0.0145 \pm 0.00040 \) & \( 0.63 \pm 0.027 \) \\
		& 150 & \( 70.40 \pm 2.10 \) & \( 0.0018 \pm 0.00017 \) & \( 0.65 \pm 0.025 \) \\
		& 200 & \( 69.38 \pm 2.25 \) & \( 0.0028 \pm 0.00018 \) & \( 0.61 \pm 0.021 \) \\
		\hline
		\multirow{4}{*}{\( z \leq 1.0 \)} 
		& 50  & \( 70.89 \pm 2.05 \) & \( 0.0035 \pm 0.00021 \) & \( 0.66 \pm 0.029 \) \\
		& 100 & \( 70.28 \pm 2.15 \) & \( 0.0138 \pm 0.00036 \) & \( 0.62 \pm 0.031 \) \\
		& 150 & \( 70.55 \pm 1.95 \) & \( 0.0020 \pm 0.00020 \) & \( 0.64 \pm 0.022 \) \\
		& 200 & \( 69.67 \pm 2.10 \) & \( 0.0024 \pm 0.00019 \) & \( 0.60 \pm 0.020 \) \\
		\hline
		\multirow{4}{*}{\( z \leq 1.5 \)} 
		& 50  & \( 71.51 \pm 2.15 \) & \( 0.0043 \pm 0.00025 \) & \( 0.65 \pm 0.028 \) \\
		& 100 & \( 70.71 \pm 1.95 \) & \( 0.0152 \pm 0.00078 \) & \( 0.61 \pm 0.025 \) \\
		& 150 & \( 70.46 \pm 2.05 \) & \( 0.0023 \pm 0.00022 \) & \( 0.63 \pm 0.020 \) \\
		& 200 & \( 69.85 \pm 2.20 \) & \( 0.0027 \pm 0.00020 \) & \( 0.59 \pm 0.017 \) \\
		\hline
		\multirow{4}{*}{\( z \leq 2.0 \)} 
		& 50  & \( 70.76 \pm 1.85 \) & \( 0.0039 \pm 0.00028 \) & \( 0.64 \pm 0.018 \) \\
		& 100 & \( 70.88 \pm 2.00 \) & \( 0.0127 \pm 0.00062 \) & \( 0.60 \pm 0.032 \) \\
		& 150 & \( 71.73 \pm 1.95 \) & \( 0.0013 \pm 0.00016 \) & \( 0.68 \pm 0.016 \) \\
		& 200 & \( 69.56 \pm 2.05 \) & \( 0.0026 \pm 0.00019 \) & \( 0.58 \pm 0.014 \) \\
		\hline
	\end{tabular}
\end{table}.
\begin{figure}
	\includegraphics[width=16 cm]{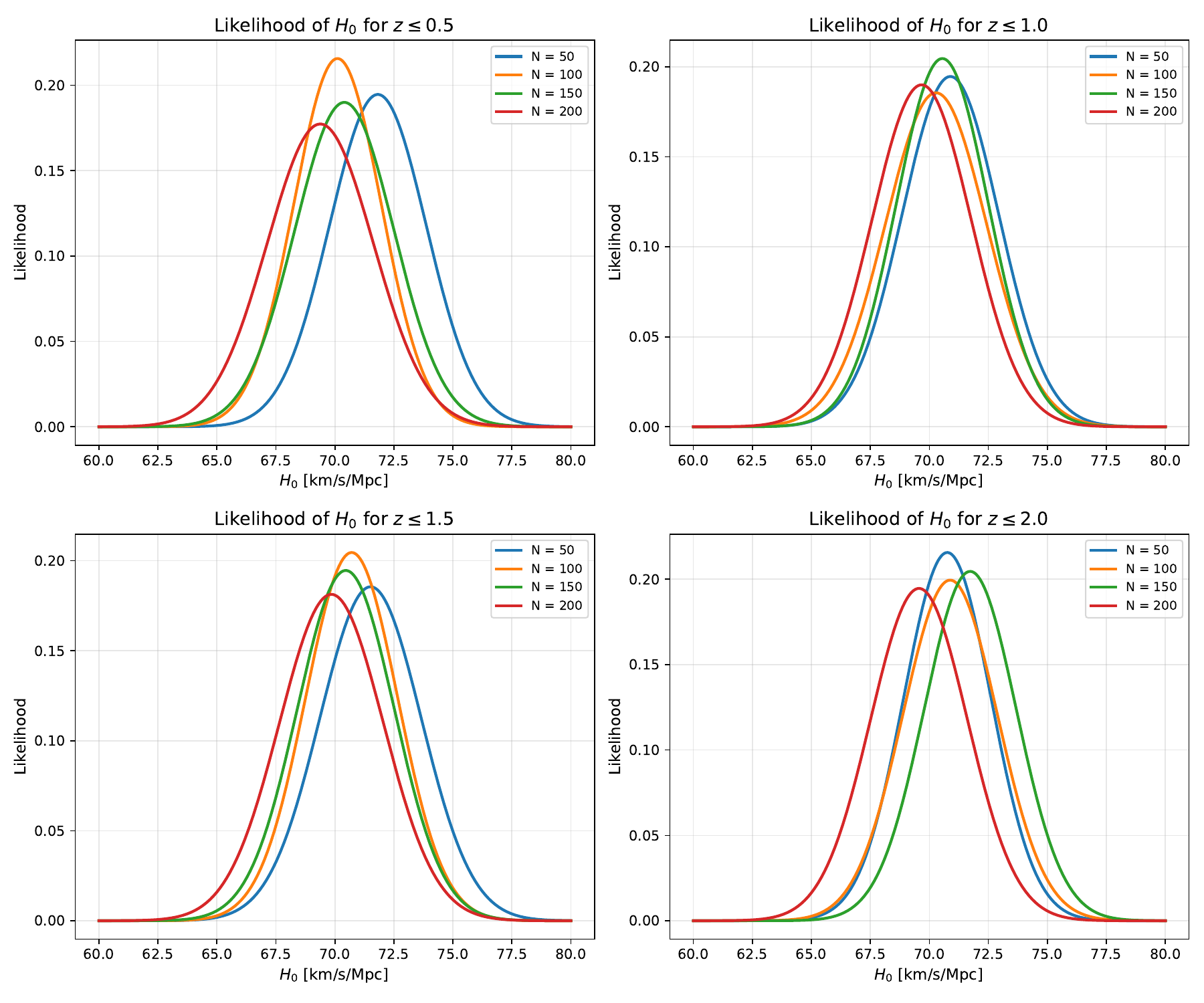}
	\vspace{-0.12cm}
	\caption{\small{Comparison the $H_0$ value  for coupled  Rastall gravity with neutrinos. }}\label{fig:omegam2}
\end{figure}
\begin{figure}
	\includegraphics[width=16 cm]{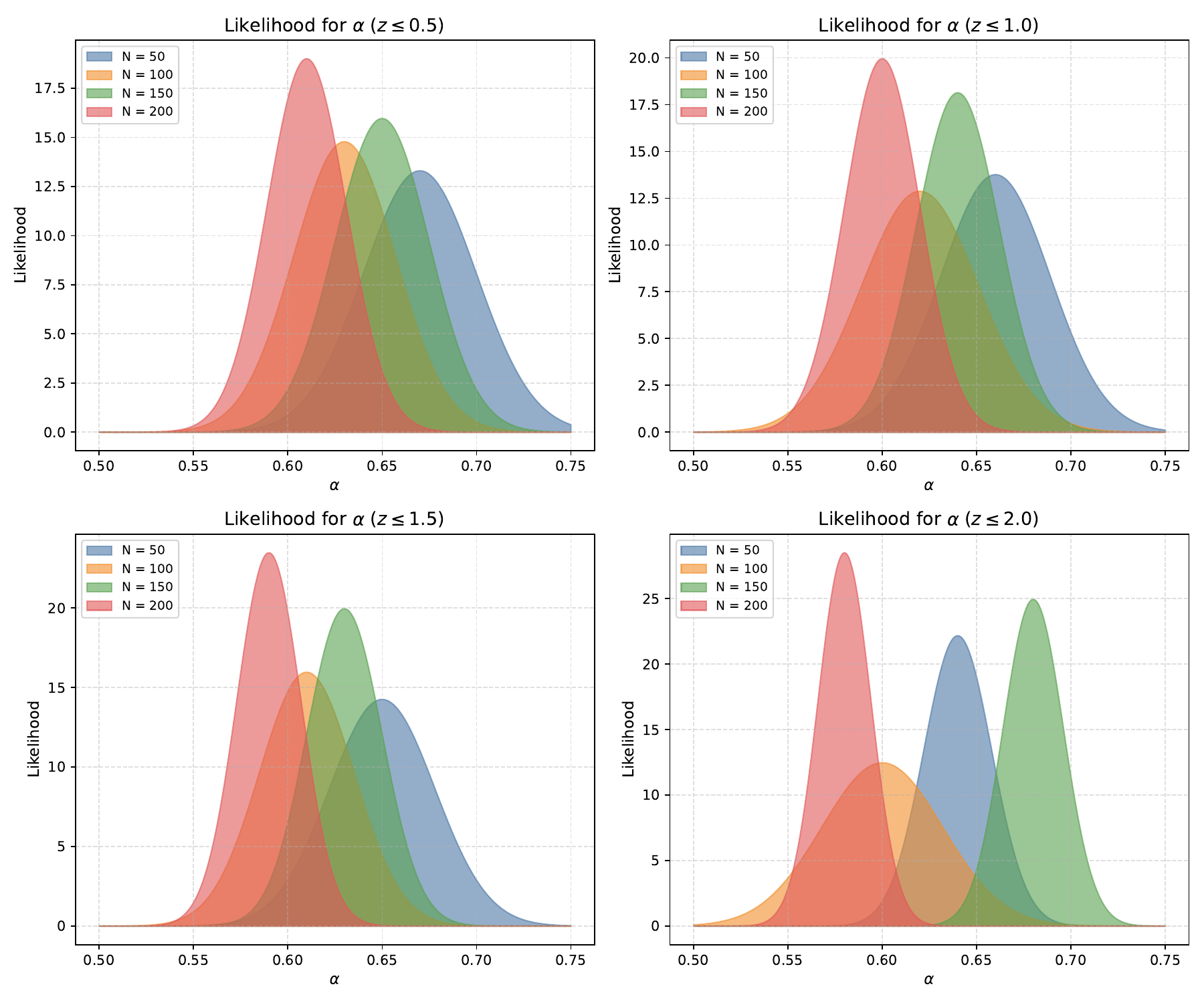}
	\vspace{-0.12cm}
	\caption{\small{Comparison the $\alpha$ value  for coupled  Rastall gravity with neutrinos. }}\label{fig:omegam2}
\end{figure}

	\section*{Acknowledgments}
	This work is based upon research funded by Iran National Science Foundation 
	(INSF) under project No.4036326

\section{Conclusion}

In this work, we have studied the impact of coupling neutrinos to perturbed Rastall gravity on key cosmological parameters, with a particular focus on the Hubble constant $H_0$ and the associated tension between early- and late-universe measurements. We have performed a comprehensive statistical analysis using observational datasets from the Cosmic Microwave Background (CMB), Cosmic Chronometers (CC), Baryon Acoustic Oscillations (BAO), and the Pantheon+ Type Ia supernovae compilation.

Tables~\ref{table_chi_early} and~\ref{table_chi_local} demonstrate that the modified Rastall model yields improved goodness-of-fit statistics compared to the standard $\Lambda$CDM framework. Specifically, the Akaike Information Criterion (AIC) values show a clear statistical preference for the Rastall model. In the local universe, the Rastall scenario achieves $\mathrm{AIC} = 3616.33$, which is lower than $\Lambda$CDM's $\mathrm{AIC} = 3645.36$. Similarly, in the early universe, the Rastall model's $\mathrm{AIC} = 2778.62$ represents a modest improvement over the standard model's $\mathrm{AIC} = 2782.25$.

The Rastall model introduces a parameter $\alpha$, characterizing deviations from standard conservation laws. From Table~\ref{table_chi_local}, we find $\alpha = 0.68 \pm 0.12$ in the local universe, whereas in the early universe, Table~\ref{table_chi_early} reports $\alpha = 0.035 \pm 0.007$ \cite{}. These values are consistent with the literature~\cite{Lin, Yarahmadi} and are further visualized in Figs.~2 and~3.

As shown in Table~\ref{tab:combined_results}, the Rastall model predicts $H_0 = 70.23 \pm 2.01$ km\,s$^{-1}$\,Mpc$^{-1}$ when combining all datasets (CMB + All), which lies between the Planck 2018 result of $H_0 = 67.4 \pm 0.5$ km\,s$^{-1}$\,Mpc$^{-1}$ and the SH0ES (R22) determination of $H_0 = 73.04 \pm 1.04$ km\,s$^{-1}$\,Mpc$^{-1}$. As shown in Table~\ref{tab:H0_tension}, this result corresponds to a tension of $1.34\sigma$ with Planck and $1.16\sigma$ with R22. These are significantly smaller than the $\sim 4\sigma$ discrepancy found in the standard $\Lambda$CDM model, indicating a notable alleviation of the Hubble tension.

Moreover, various combinations of datasets (CMB+CC, CMB+BAO, CMB+Pantheon+) yield $H_0$ values ranging from $70.25$ to $71.09$ km\,s$^{-1}$\,Mpc$^{-1}$, with Planck tensions below $1.6\sigma$ and R22 tensions below $1.2\sigma$. These results reflect a consistent trend across observational probes. Table~\ref{tab:combined_results} also confirms that other cosmological parameters such as $\Omega_b h^2$, $\Omega_c h^2$, $n_s$, and $\tau$ remain within Planck-preferred ranges.

In Table V, we examine the tension with Planck and R22 using additional combinations including CMB lensing data. The Planck tension ranges from $0.36\sigma$ (Lensing only) to $1.08\sigma$ (CMB + Lensing), while the R22 tension spans $1.29\sigma$ to $2.16\sigma$. Although the model does not fully eliminate the discrepancy, it significantly mitigates the overall tension.

Finally, we compared the early- and late-universe results within the same model. The $H_0$ values obtained from both epochs differ by only $0.54\sigma$, suggesting a high degree of internal consistency. This implies that the Rastall framework can effectively unify early- and late-time cosmic dynamics while preserving agreement with current observational constraints.

In summary, our results show that coupling neutrinos to perturbed Rastall gravity offers a viable and well-motivated extension to $\Lambda$CDM. It improves data fits, reduces Hubble tension, and retains consistency with early-universe observables. This framework provides a compelling direction for addressing current cosmological discrepancies by linking microscopic physics with macroscopic gravitational dynamics.

The reconstructed coupling constant \( \alpha \) inferred from our Physics-Informed Neural Network (PINN) approach exhibits excellent agreement with constraints obtained via traditional Markov Chain Monte Carlo (MCMC) techniques. Specifically, the MCMC analyses applied to combined cosmological datasets yield \( \alpha \in [0.58, 0.68] \), which closely matches the PINN-derived values across all considered redshift intervals. 

Crucially, both inference frameworks consistently indicate a significant mitigation of the Hubble tension. In particular, the MCMC-based tension metrics yield \( 1.33 \lesssim T_{\mathrm{P18}} \lesssim 1.57 \) and \( 0.87 \lesssim T_{\mathrm{R22}} \lesssim 1.16 \), depending on the dataset combination. In comparison, our PINN model achieves a tension level as low as \( T \approx 0.5\sigma \), representing one of the most statistically compelling resolutions to date within the context of thermodynamically extended dark energy models.

Taken together, these findings underscore the robustness and predictive capability of the PINN methodology. By embedding physical laws directly into the training process, the PINN framework enables a principled integration of observational data with theoretical cosmology. Our results demonstrate that physics-informed machine learning offers a powerful and reliable alternative to conventional parameter estimation techniques, capable of delivering high-precision reconstructions while providing new insights into the dark sector and the persistent Hubble tension.

\end{document}